\documentclass{article}
\usepackage[utf8]{inputenc}
\usepackage{geometry}
\usepackage{auxlib}
\usepackage{authblk}

\newcommand{\alg}{\mathrm{ALG}}
\newcommand{\opt}{\mathrm{OPT}}

\newcommand{\lp}{\mathsf{LP^*}}

\newcommand{\gs}{\mathsf{GS}}

\newcommand{\flag}{\textup{\texttt{flag}}}
\newcommand{\true}{\textup{\texttt{true}}}
\newcommand{\false}{\textup{\texttt{false}}}

\newcommand{\bT}{\mathbf{T}}
\newcommand{\bV}{\mathbf{V}}

\newcommand{\bE}{\mathbf{E}}

\newcommand{\tw}{\mathsf{tw}}

\newcommand{\p}{\textup{\texttt{1}}}
\newcommand{\n}{\textup{\texttt{0}}}
\newcommand{\X}{\mathsf{X}}

\newcommand{\chng}{\textup{\texttt{chng}}}
\newcommand{\conn}{\textup{\texttt{conn}}}

\newcommand{\inedge}{\delta_{\text{in}}}
\newcommand{\outedge}{\delta_{\text{out}}}

\newcommand{\edge}{\mathsf{edge}}

\newcommand{\lbl}{\mathsf{label}}
\newcommand{\tto}{\text{-}}
\newcommand{\leaf}{\mathsf{leaf}}

\newcommand{\kid}{\mathsf{child}}

\newcommand{\xv}{x_{\bv}}
\newcommand{\xu}{x_{\bu}}
\newcommand{\xr}{x_{\br}}

\newcommand{\poly}{\text{poly}}

\usepackage[square,sort,comma,numbers]{natbib}

\title{Polylogarithmic Approximation for Robust s-t Path\thanks{All authors (ordered alphabetically) have equal contributions and are corresponding authors. 
}}
\author[1]{Shi Li}
\author[2]{Chenyang Xu}
\author[3]{Ruilong Zhang}

\affil[1]{\small Department of Computer Science and Technology, Nanjing University}
\affil[2]{\small Software Engineering Institute, East China Normal University}
\affil[3]{\small Department of Computer Science and Engineering, University at Buffalo}

\affil[ ]{\texttt{shili@nju.edu.cn, cyxu@sei.ecnu.edu.cn, ruilongz@buffalo.edu}}
\date{}

\begin{document}

\maketitle

\begin{abstract}
The paper revisits the robust $s$-$t$ path problem, one of the most fundamental problems in robust optimization. In the problem, we are given a directed graph with $n$ vertices and $k$ distinct cost functions (scenarios) defined over edges, and aim to choose an $s$-$t$ path such that the total cost of the path is always provable no matter which scenario is realized. 
With the view of each cost function being associated with an agent, our goal is to find a common $s$-$t$ path minimizing the maximum objective among all agents, and thus create a fair solution for them.
The problem is hard to approximate within $o(\log k)$ by any quasi-polynomial time algorithm unless $\NP \subseteq \mathrm{DTIME}(n^{\poly\log n})$, and the best approximation ratio known to date is $\widetilde{O}(\sqrt{n})$ which is based on the natural flow linear program. 
A longstanding open question is whether we can achieve a polylogarithmic approximation even when a quasi-polynomial running time is allowed. 

Our main result is a $O(\log n \log k)$ approximation for the Robust s-t Path in quasipolynomial time, solving the open question in the quasi-polynomial time regime.
The algorithm is built on a novel linear program formulation for a decision-tree-type structure,
which enables us to get rid of the $\Omega(\sqrt{n})$ integrality gap for the natural flow LP.
Further, we also consider some well-known graph classes, e.g., graphs with bounded treewidth, and show that the polylogarithmic approximation can be achieved polynomially on these graphs. 
We hope the new proposed techniques in the paper can offer new insights into the robust $s$-$t$ path problem and related problems in robust optimization.
\end{abstract}

\newpage


\section{Introduction}
\label{sec:intro}

Routing optimization under uncertainty is one of the most important and challenging computational tasks in the real world. In many scenarios, the uncertainty of the network is frequently encountered and must be dealt with. For example, a traveler wants to minimize the total travel time in a transportation network, but the travel time of each road segment is often uncertain due to accidents and traffic jams. 
There are two classical paradigms used to tackle this issue: stochastic optimzation~\cite{DBLP:conf/nips/ChenYL22,DBLP:conf/nips/CutlerDH21,DBLP:conf/icml/0002CMK20,DBLP:books/daglib/0017615} and robust optimzation~\cite{DBLP:conf/nips/AnG21,DBLP:books/degruyter/Ben-TalGN09,DBLP:journals/eor/GabrelMT14,DBLP:conf/nips/LiG21}.
In stochastic optimization, we are given a pre-specified distribution of the uncertain information and aim to optimize the average performance over the distribution; while for robust optimization, the worst-case performance over the distribution is the target.

The paper focuses on robust routing optimization and revisits the robust $s$-$t$ path problem~\cite{DBLP:journals/cor/GangJ98}, one of the most fundamental problems in this area. 
In the problem, there are several edge cost functions for a given graph and the goal is to find an $s$-$t$ path that minimizes the maximum cost across all the edge cost functions. 
We notice that besides viewing this problem as an optimization model under uncertainty, it can also be interpreted as a multi-objective optimization problem or a fairness computation problem among multiple agents.
In a routing network, a link (edge) usually has several attributes, e.g., the usage cost, the delay, and so on. Denoting each attribute by a cost function, we can formulate the multi-objective routing problem~\cite{DBLP:journals/eor/DuqueLM15,DBLP:books/daglib/0022057,DBLP:journals/eor/Sedeno-NodaC19} as our model.
In the context of fairness computation, different edge cost functions can be viewed as different perspectives of agents on the edges, and our problem aims to find a public path that can satisfy them under the min-max fairness notion~\cite{DBLP:conf/icml/AbernethyAKM0Z22,DBLP:conf/icml/MartinezBS20,DBLP:journals/ton/RadunovicB07}.

The robust $s\tto t$ path problem was first studied by~\cite{DBLP:journals/cor/GangJ98}, and since then it has received widespread attention due to its broad applicability.
In~\cite{DBLP:journals/cor/GangJ98}, the authors show that the problem is strongly 
NP-Hard even if there are only two scenarios.
Later, \cite{DBLP:journals/eor/AissiBV07} considers the problem with the constant number of scenarios, and they show that the problem admits a fully polynomial-time approximate scheme (FPTAS).
For the setting that the number of scenarios is part of the input, it is easy to see that simply computing the shortest path w.r.t the summation of the $k$ cost functions can obtain an approximation ratio of $O(k)$. 
Kasperski and Zielinski~\cite{DBLP:journals/ipl/KasperskiZ09} prove that the problem is hard to approximate within $o(\log k)$ unless $\NP \subseteq \text{DTIME}(n^{\poly\log n})$.
It has been open ever since whether a polylogarithmic approximation can be achieved.
A recent breakthrough in the approximation ratio is made by Kasperski and Zielinski~\cite{DBLP:journals/corr/abs-1806-08936} in which they gave a flow-LP-based algorithm that is $\widetilde{O}(\sqrt{n})$\footnote{We use $\widetilde{O}$ to hide a polylogarithmic factor.}-approximate. 
They further showed that their analysis is nearly tight by proving an integrality gap of $\Omega(\sqrt{n})$ for the flow-based LP. 

It should be noted that Bilò et al.~\cite{DBLP:conf/icalp/BiloC0FM17} studied the $\ell_q$-norm shortest path problem which is a generalized version of robust $s$-$t$ path, i.e., the problem aims to find a $s$-$t$ path $\cP$ to minimize the value of $\left( \sum_{i\in[k]}c_i(\cP)^q \right)^{1/q}$, where $c_i(\cP)$ is agent $i$'s cost for the selected path $\cP$.
Their algorithm~\cite{DBLP:conf/icalp/BiloC0FM17} (Algorithm 2) extends the classical Dijkstra algorithm by replacing the distance with the $\ell_q$-norm metric.
Our directed graph is a DAG, so the Dijkstra-type algorithm becomes a dynamic programming-type algorithm, with nodes processed using a topological order.
So, their algorithm just stores the best path in the $\ell_q$-norm for every node.
It is claimed in~\cite{DBLP:conf/icalp/BiloC0FM17} (Theorem 14) that such an algorithm achieves $O(\min\{q,\log k\})$-approximation for the $\ell_q$-norm shortest path problem.
However, unfortunately, there exists a crucial error in the analysis. 
In \cref{app:hard-greedy}, we give a hard instance on a series-parallel graph for the algorithm and show that the approximation ratio is at least $k^{1-1/q}$.
In other words, when $q=O(\log k)$, the proposed algorithm~\cite{DBLP:conf/icalp/BiloC0FM17}(Algorithm 2) is $O(k)$-approximate for the robust $s$-$t$ problem.

\subsection{Our Contributions}

In the paper, we make significant progress in closing the gap between the known upper and the lower bound for robust $s\tto t$ path. 
We show that for two natural graph classes, the polylogarithmic approximation can be obtained in polynomial time; while for general graphs, there exists a polylogarithmic approximated algorithm running in a quasi-polynomial time. 
In the following, we first give the formal definition of our problem, and then summarize the main results.

\paragraph{ The Robust $s\tto t$ Path Problem.}
Consider a directed graph $G(V, E)$ with $n$ vertices and $m$ edges. There are $k$ scenarios (which are also referred to as agents in the following), where each scenario $i\in [k]$ has an edge cost function $c_i: 2^{E}\to \R_{\geq 0}$. Given two specified vertices $s$ and $t$ in the graph, the goal is to find an $s\tto t$ path $\cP$ such that $\max_{i\in[k]}c_i(\cP)$ is minimized, where $c_i(\cP):=\sum_{e\in\cP} c_i(e)$.

\paragraph{ Main Result 1 (\cref{thm:sp-graph}).}
Given any series-parallel graph, there is a randomized algorithm that can achieve an approximation ratio of $O(H\log k)$ in polynomial time, where $H$ is the height of 
the decomposition tree of the series-parallel graph and $k$ is the number of agents.

\medskip

Our first result is on a specific category of graphs called {\em series-parallel graphs} (\cref{sec:sp}), which are used by ~\cite{DBLP:journals/ipl/KasperskiZ09} to demonstrate a lower bound of $\Omega(\log^{1-\epsilon}k)$ (for any $\epsilon>0$) for the robust $s\tto t$ path problem. 
We begin by showing that the natural flow linear program (LP) on a series-parallel graph has an integrality gap of $\Omega(k)$ and $\Omega(\sqrt{n})$ (\cref{sec:flow-lp-gap}).
The gaps hold even when we integrate the knowledge of the optimum cost to the LP to circumvent some obvious gap instances.
This result aligns with the prior findings of~\cite{DBLP:journals/corr/abs-1806-08936}, but our constructed instance is significantly simpler. 
It should be noted that most prior algorithms in the existing literature rely on the flow LP mentioned above, and thus, their approximation ratios cannot be better than $O(\min\{k,\sqrt{n}\})$.

To overcome the pessimistic gap, we develop a novel linear program based on the {\em decomposition tree} of the series-parallel graph. 
Subsequently, we demonstrate that a dependent randomized rounding algorithm for the LP can obtain an approximation ratio of $O(H \log k)$.
Particularly, for the hard instance that leads to a lower bound $\Omega(\log^{1-\epsilon}k)$ (for any $\epsilon>0$) stated in~\cite{DBLP:journals/ipl/KasperskiZ09}, our algorithm can return a $O(\log \log n \log k)$-approximate solution, which is nearly tight since there is only a $O(\log \log n)$-gap
(see \cref{sec:sp:hard-instance} for details).

\paragraph{ Main Result 2 (\cref{thm:general:graph}, \ref{thm:quasi:hardness}).}
Given any directed graph, there is a randomized algorithm that can obtain a $O(\log n \log k)$-approximate solution in quasi-polynomial time, where $n$ is the number of vertices and $k$ is the number of agents.
Moreover, any quasi-polynomial time algorithm for robust $s$-$t$ path has an approximation lower bound of $\Omega(\log^{1-\epsilon} k)$ (even on series-parallel graphs) under the assumption that $\NP\nsubseteq \mathrm{DTIME}(n^{\poly \log n})$.

\medskip

Finally, we consider general graphs (\cref{sec:general-graphs}). 
The algorithm is also LP-based, following a similar framework as the algorithm for series-parallel graphs.
The main challenge here is that we no longer have a simple tree structure for general graphs.
To address this issue, we construct a decision-tree-type tree structure for the given graph and write a linear program based on it.
Our algorithm then builds on this new LP to give the first polylogarithmic approximation for general graphs.
Additionally, we show that the lower bound of $\Omega(\log^{1-\epsilon} k)$ can be extended to the algorithms running in quasi-polynomial time, i.e., the problem is still hard to be approximated within $o(\log k)$ even if we allow quasi-polynomial time algorithms (\cref{sec:quasi:hardness}).

\paragraph{ Main Result 3 (\cref{thm:ratio:treewidth}).} 
Given any graph with bounded treewidth, there is a polynomial time algorithm with an approximation ratio of $O(\log n \log k)$, where $n$ is the number of vertices and $k$ is the number of agents.

\medskip

We then consider a more general graph class called {\em graphs with bounded treewidth} (\cref{sec:treewidth}). 
This graph class admits the class of series-parallel graphs as its special case that the treewidth $\tw=2$, and therefore, combining the above two results gives a $O(\min\{H,\log n\}\cdot\log k)$-approximation for series-parallel graphs.
Besides series-parallel graphs, the graph class includes many other common graphs, such as trees ($\tw=1$), pseudoforests ($\tw=2$), Cactus graphs ($\tw=2$), outerplanar graphs ($\tw=3$), Halin graphs ($\tw=3$), and so on.
In this part, we employ the nice properties provided by the treewidth decomposition of these graphs and obtain a polylogarithmic approximation.
This result improves upon the previous state-of-the-art ratio of $\widetilde{O}(\sqrt{n})$~\cite{DBLP:journals/corr/abs-1806-08936}.

\paragraph{Main Result 4 (\cref{thm:hardness:maximin-st}, \ref{thm:hardness:maximin-is}, \ref{thm:hardness:maximin-stree}).}
For the problems of robust $s\tto t$ path, weighted independent set, and spanning tree under the \emph{maximin criteria}, it is NP-Hard to determine whether their instances have zero-cost optimal solutions or not. 
This implies that these problems do not admit any polynomial time $\alpha$-approximate algorithm unless $\PP=\NP$, where $\alpha$ is an arbitrary function of the input.

\medskip

The paper also considers the \emph{maximin criteria}, where the goal is to maximize the minimum cost among all agents (\cref{sec:hardness}).
By observing that the classic algorithms (e.g., Dijkstra's algorithm) that work for the shortest path problem on DAGs (directed acyclic graphs) also work for the longest path problem on DAGs, one might expect that the maximin criteria is also a candidate objective to investigate the robustness of the $s\tto t$ path problem, i.e., finding an $s\tto t$ path $\cP$ such that $\min_{i\in[k]}c_i(\cP)$ is maximized.
We demonstrate this is not the case by providing a strong lower bound for the problem under the maximin criteria.
Our reduction builds on a variant of the set cover problem. 
Employing a similar basic idea of the reduction, we also show that the maximin weighted independent set problem on trees or interval graphs is not approximable.
This constitutes a strong lower bound for this problem, while the previous works~\cite{DBLP:journals/symmetry/KlobucarM21,DBLP:journals/ol/NobibonL14} only show the NP-Hardness.
Our reduction idea can further be extended to the maximin spanning tree problem, which implies that the robust spanning tree problem is also not approximable under the maximin objective.

\subsection{Other Related Works}

\paragraph{Robust Minimax Combinatorial Optimization.}

Robust minimax optimization under different combinatorial structures has been extensively studied in the past three decades.
See~\cite{aissi2009min,kasperski2016robust} for a survey.
All these problems are shown to be NP-Hard even for simple constraints such as spanning trees, knapsacks, $s \tto t$ cuts, and perfect matching on bipartite graphs~\cite{kasperski2016robust}.
Besides these fundamental constraints, the minimax submodular ranking problem was studied in~\cite{DBLP:journals/corr/abs-2212-07682} very recently.
For minimax spanning tree, a $O(\log k/\log \log k)$-approximation algorithm is known~\cite{DBLP:conf/focs/ChekuriVZ10}, which is almost tight by the lower bound of $\Omega(\log^{1-\epsilon} k)$ stated in~\cite{DBLP:journals/tcs/KasperskiZ11}.
The problem of minimax perfect matching has a lower bound of $\Omega(\log^{1-\epsilon} k)$~\cite{DBLP:journals/ipl/KasperskiZ09}, while the best upper bound so far is still $O(k)$ which is trivial.
In the case where $k$ is a constant, fully polynomial time approximation schemes are known for spanning trees, perfect matching, knapsacks, and $s \tto t$ paths~\cite{DBLP:journals/disopt/AissiBV10,aissi2009min,kasperski2016robust,DBLP:conf/focs/PapadimitriouY00}.

\paragraph{Multiobjective $s \tto t$ Path.}

Finding an $s \tto t$ path is a fundamental problem in multi-objective optimization~\cite{DBLP:books/daglib/0022057}.
An Excellent survey of multiobjective combinatorial optimization, including multiobjective $s \tto t$ path, can be found in~\cite{DBLP:journals/anor/ChinchuluunP07}.
Typically, we are given a directed graph $G:=(V,E)$.
Each edge $e\in E$ has a positive cost vector $\bc(e):=(c_1(e),\ldots,c_k(e))$.
For every $s \tto t$ path $\cP\subseteq E$, we have a cost vector $\bc(\cP)=(c_1(\cP),\ldots,c_k(\cP))$ with $c_i(\cP)=\sum_{e\in \cP} c_i(e)$.
The goal is to compute an $s \tto t$ path $\cP$ such that $\cP$ is Pareto optimal.
Not surprisingly, this problem has been shown to be NP-hard even if the cost vector only has two coordinates~\cite{serafini1987some} in which the problem is called the objective $s \tto t$ path minimization problem.
Biobjective $s \tto t$ path minimization has also been studied extensively~\cite{DBLP:journals/eor/DuqueLM15,DBLP:journals/eor/Sedeno-NodaC19}, in which researchers mainly focus on the exact algorithms with exponential running time.
In addition, a fully polynomial-time approximation scheme (FPTAS) is proposed by~\cite{DBLP:conf/focs/PapadimitriouY00}.

\paragraph{Fair Allocation with Public Goods.}
By observing the minimax objective as a fairness criterion, our problem shares some similarities with the problem of public goods, which was first used to distinguish the previous private goods by Conitzer et al.~\cite{DBLP:conf/sigecom/ConitzerF017} in the field of fair division.
Specifically, there is a multiagent system and different agents hold different options for the same goods.
And, they aim to select a feasible set of goods to satisfy the various fairness notions, such as propositional share or its generalization~\cite{DBLP:conf/sigecom/ConitzerF017,DBLP:conf/sigecom/FainM018}.
In~\cite{DBLP:conf/sigecom/FainM018}, they study some constraints of goods, i.e., the selected goods must form a matching or matroid.
The minimax criterion is quite different from other fairness measures in the fair division field, which leads to different techniques.

\subsection{Roadmap}

We first show in \cref{sec:flow-lp-gap} that the natural min-cost-flow-based linear program has an integrality gap of $\Omega(k)$ and $\Omega(\sqrt{n}\})$.
To enhance the paper's readability, we shall first present the algorithm for series-parallel graphs in \cref{sec:sp}.
Then, we give the algorithm for general graphs in \cref{sec:general-graphs}.
We then describe the algorithm for graphs with bounded treewidth in \cref{sec:treewidth}.
Finally, we show in \cref{sec:hardness} that several fundamental robust problems are not approximable under the maximin criteria.

\section{The Integrality Gap of Flow-LP}
\label{sec:flow-lp-gap}

In this section, we show that the natural flow LP has a pessimistic integrality gap. The program is as follows:

\begin{align*}
    \text{min} && T & \tag{\text{Flow-LP}} \label{Flow-LP}\\
    \text{s.t.} \nonumber \\
    &&\sum_{e\in E}x_e \cdot c_i(e) &\leq T, &\forall i\in [k] \\
    &&\sum_{e\in\inedge(v)}x_e &= \sum_{e\in\outedge(v)}x_e, &\forall v\in V\setminus\set{s,t} \\
    &&\sum_{e\in\outedge(s)}x_e &= \sum_{e\in\inedge(t)} x_e =1, & \\
    &&x_e &\geq 0, &\forall e\in E 
\end{align*}

An integrality gap of $\Omega(\max\{k,n\})$ can be shown by the following simple hard instance.
Consider the directed graph $G:=(V,E)$ such that (\rom{1}) there are only two vertices, i.e., $V:=\set{s,t}$; (\rom{2}) there are $k$ edges from $s$ to $t$, i.e., $E:=\set{e_1,\ldots,e_k}$\footnote{Note that although the current graph contains multiple edges, we can easily convert it into a simple graph by adding $k+1$ dummy vertices between $s$ and $t$. Thus, we directly assume that the number of vertices $n = k+2$.}.
For each agent $i\in [k]$, the cost function is defined as follows: (\rom{1}) $c_i(e_i)=1$; (\rom{2}) $c_i(e_j)=0$ for any $j\notin \{i,k+1\}$.
The optimal integral solution will pick an arbitrary edge $e_{i}$ and thus has an objective value of $1$, while the optimal fractional solution is at most $\frac{1}{k}$ by setting $x_{e_i}=\frac{1}{k}$ for each $i\in [k]$.
Thus, this gives us a gap of $k$ or $n-2$.

One possible way to enhance \eqref{Flow-LP} is to integrate the truncated instance and doubling trick technique initially proposed by~\cite{DBLP:journals/mp/LenstraST90}. 
Specifically, we can apply the doubling technique to guess the optimal value. 
For every trial $\hat{T}$, we start by truncating the instance through the elimination of some invalid edges: for an edge $e$, if there is an agent $i$ such that $c_i(e)>\hat{T}$, delete this edge from the graph. 
We then employ linear programming to verify the existence of a feasible solution to the truncated instance. 
To formalize this, we introduce the following refined LP. Note that this program does not have an objective, and we only leverage it to check the feasibility of the guessed $\hat{T}$.

\begin{align}
    &&  & \tag{\text{Enh-Flow-LP}} \label{En-Flow-LP}\\
    &&\sum_{e\in E}x_e \cdot c_i(e) &\leq \hat{T},  &\forall i\in [k] \label{flow-LPC:cost}\\
    &&\sum_{e\in\inedge(v)}x_e &= \sum_{e\in\outedge(v)}x_e, &\forall v\in V\setminus\set{s,t} \label{flow-LPC:st}\\
    &&\sum_{e\in\outedge(s)}x_e &= \sum_{e\in\inedge(t)} x_e =1, & \label{flow-LPC:equal} \\
    &&x_e &=0, &\text{if } \exists i\in[k] \text{ s.t. } c_i(e)>\hat{T} \\
    &&x_e &\geq 0, &\forall e\in E 
\end{align}

\begin{figure}[tb]
    \centering
    \includegraphics[width=13cm]{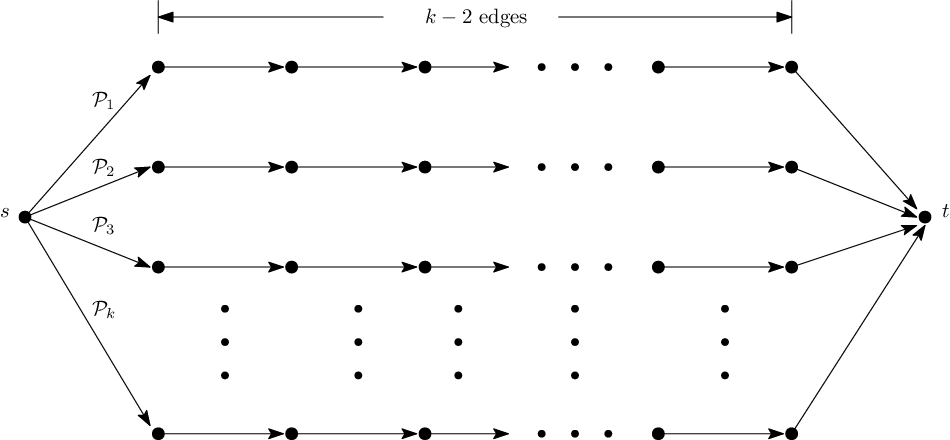}
    \caption{Hard instance for \eqref{En-Flow-LP}. The optimal integral solution must take one of $k$ paths while the optimal fractional solution takes each path by $\frac{1}{k}$.}
    \label{fig:flow_lp_gap}
\end{figure}

\eqref{En-Flow-LP} eliminates the above hard instance since all edges will be discarded when we set $\hat{T}<1-\epsilon$.
Hence, \eqref{Flow-LP} has no feasible solution if $\hat{T}<1-\epsilon$.
However, even with \eqref{En-Flow-LP}, we are still not able to get rid of an integrality gap of $\Omega(k)$. 
The hard instance is shown in \cref{fig:flow_lp_gap}.
The directed graph $G:=(V,E)$ is composed of $k$ disjoint $s\tto t$ paths, denoted as $\cP_1,\ldots,\cP_k$. 
For each agent $i$, the cost function is defined as follows: (i) $c_i(e)=1$ for all $e\in\cP_i$; (ii) $c_i(e)=0$ for all edges in $E\setminus \cP_i$. 
Thus, $c_i(\cP_i)=k$ and $c_i(\cP_j)=0$ for all $j\ne i$. 
The optimal integral solution requires the selection of one of the paths, yielding an objective value of $k$. 
However, the optimal fractional solution would take all paths by $\frac{1}{k}$, resulting in an objective value of $1$, since no edge is dropped even when $\hat{T}$ is set to $1$. 
Therefore, \eqref{En-Flow-LP} still has an integrality gap of $\Omega(k)$. 
This hard instance also implies that \eqref{En-Flow-LP} has an integrality gap of $\Omega(\sqrt{n})$, where $n$ is the number of vertices. 
This result matches the finding in~\cite{DBLP:journals/corr/abs-1806-08936}, but our proof is simpler.

\section{Series-Parallel Graphs}
\label{sec:sp}

In this section, we show that there is a randomized algorithm that achieves $O(H \log k)$ approximation for series-parallel graphs, where $H$ is the height of the series-parallel graph's decomposition tree; its meaning will be clear later. 
The algorithm can be viewed as a warm-up example for the general graph, as the algorithm for the general graph follows a similar algorithmic framework.
The algorithm for the general graph follows a similar algorithmic framework and it is able to achieve the same approximation ratio, but we have to solve a linear program of quasi-polynomial size which leads to a quasi-polynomial time algorithm.
Formally, we shall show the following theorem (\cref{thm:sp-graph}) in this section.

\begin{theorem}
Given any series-parallel graph $G$, there is a polynomial time algorithm that returns a $O(H \log k)$-approximation solution with probability at least $1-(\frac{1}{k}+\frac{1}{kH})$ for robust $s\tto t$ path, where $H$ is the height of $G$'s decomposition tree and $k$ is the number of agents.
\label{thm:sp-graph}
\end{theorem}

In \cref{sec:sp-graph:concepts}, we give the basic concepts and properties of the series-parallel graphs, which we will use later to build our linear programming formulation.
In \cref{sec:sp-graph:lp}, we formally present our LP formulation, and we give a rounding algorithm and its analysis in \cref{sec:sp-graph:alg}.

\subsection{Basic Concepts}
\label{sec:sp-graph:concepts}

\begin{definition}[Series-Parallel Graph]
A directed graph $G:=(V,E,s,t)$ with source $s$ and sink $t$ is called a {\em series-parallel graph}, if it contains a single edge from $s$ to $t$, or it can be built inductively using the following {\em series} and {\em parallel composition} operations.
The series composition of two-terminal graphs $G_1:=(V_1,E_1,s_1,t_1)$ and $G_2:=(V_2,E_2,s_2,t_2)$ is to identify $t_1$ and $s_2$, and let $s_1$ and $t_2$ be the new source and sink in the resulting graph. 
The parallel composition of two-terminal graphs $G_1:=(V_1,E_1,s_1,t_1)$ and $G_2:=(V_2,E_2,s_2,t_2)$ is to identify $s_1$ with $s_2$ and $t_1$ with $t_2$ respectively, and let $s_1 = s_2$ and $t_1 = t_2$ be the new source and sink.
\label{def:sp-graph}
\end{definition}

\begin{figure}[htb]
\centering
\includegraphics[width=15cm]{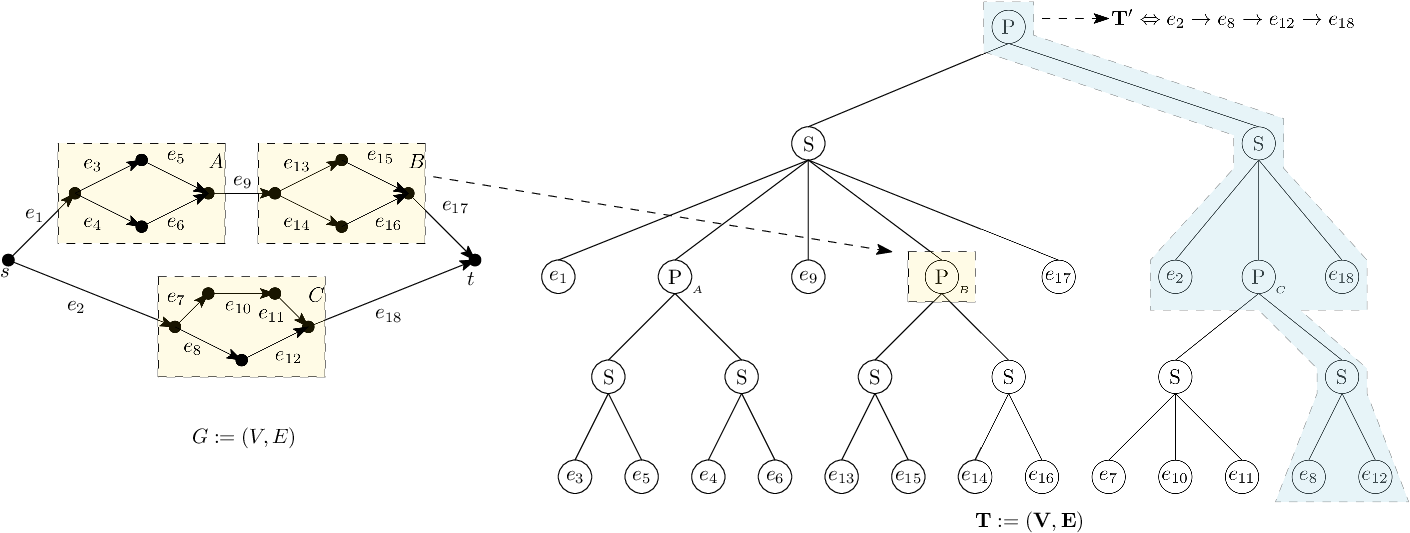}
    \caption{An example of the decomposition tree of a series-parallel graph. The series-parallel graph $G:=(V,E)$ is shown on the left and its decomposition tree $\bT:=(\bV,\bE)$ is shown on the right. Each leaf node in $\bT$ corresponds to an edge in $E$. Each internal node is either a series node S or a parallel node P. And it indicates how to merge the child nodes' subgraph. For example, consider the S node and its two child nodes $e_{13}$ and $e_{15}$. Then, the subgraph of this S node is $e_{13}\to e_{15}$ which merges its two child nodes' subgraph via the series composition. And also, the subgraph of this S node's parent corresponds to the subgraph $B$ in $G$, which merges $e_{13}\to e_{15}$ and $e_{14}\to e_{16}$ via the parallel composition. An $s \tto t$ path corresponds to a feasible subtree (\cref{def:sp:feasible-subtree}). For example, the feasible subtree $\bT'$ can be converted to an $s \tto t$ path $e_{2}\to e_{8} \to e_{12} \to e_{18}$.}
    \label{fig:sp-tree}
\end{figure}

A series-parallel graph can be represented in a natural way by a tree structure that describes how to assemble some small graphs into a final series-parallel graph through series and parallel composition. 
Such a tree structure is commonly called the {\em decomposition tree} of the series-parallel graph in the literature~\cite{DBLP:journals/siamcomp/ValdesTL82}.
Formally, a decomposition tree $\bT:=(\bV,\bE)$ of a series-parallel graph $G:=(V,E)$ is a tree such that 
(\rom{1}) each leaf node $\bu\in\bV$ corresponds an edge in $E$; 
(\rom{2}) each internal node is either a series or parallel node;
(\rom{3}) the child nodes of a parallel (resp. series) node must be leaf nodes or series (resp. parallel) nodes.
The series (resp. parallel) node corresponds to the series (resp. parallel) composition, and they are used to indicate how to merge the small subgraphs of its child nodes. 
The subgraph $G_{\bu}$ of a node $\bu$ is a subgraph of $G$ such that $G_{\bu}$ only contains those edges corresponding to the leaf nodes in the subtree rooted at $\bu$.
Let $H$ be the height of $\bT$.
Given a series-parallel graph, it is known that its decomposition tree can be built in linear time by the standard series-parallel graph recognition algorithm~\cite{DBLP:journals/siamcomp/ValdesTL82}.
An example can be found in \cref{fig:sp-tree}.

We remark that the children of a parallel node are unordered, and for a series node, the children should be considered as ordered. However, for the s-t path problem, the order is not important as permuting the children of a series node will lead to an equivalent instance.

We aim to give a linear program based on the decomposition tree $\bT$.
Clearly, not all subtrees of $\bT$ correspond to an $s \tto t$ path of $G$.
In the following, we introduce the concept of a special subtree of $\bT$ called {\em feasible subtree} (\cref{def:sp:feasible-subtree}), which is able to be converted into an $s \tto t$ path.

\begin{definition}[Feasible Subtree]
A subtree $\bT'$ of $\bT$ is called a feasible subtree if and only if 
(\rom{1}) $\bT'$ includes the root node of $\bT$;
(\rom{2}) for every series node $\bs$ in $\bT'$, $\bT'$ includes all child nodes of $\bs$;
(\rom{3}) for every parallel node $\bp$ in $\bT'$, $\bT'$ includes exactly one child node of $\bp$.
\label{def:sp:feasible-subtree}
\end{definition}

It is not hard to see that there is a one-to-one correspondence between the feasible subtree of $\bT$ and the $s \tto t$ path of $G$.
Recall that we aim to give a linear program based on the decomposition tree.
Thus, it now remains to define a cost function $f_i:2^{\bV}\to\R_{\geq 0}$ according to the cost function $c_i$.
Since each edge corresponds to a unique leaf node in $\bT$, it is easy to define $f_i$ by $c_i$: 
for each $\bv\in\bV$, $f_i(\bv):=c_i(e)$ if node $\bv$ corresponds to some edge $e$; otherwise $f_i(\bv):=0$.
Formally, we have the following simple observation (\cref{obs:sp-graph:subtree=stpath}).

\begin{observation}
Given any series-parallel graph $G:=(V,E)$ and its decomposition tree $\bT:=(\bV,\bE)$, fix an arbitrary agent $i\in[k]$, any $s \tto t$ path $\cP$ of $G$ with the cost $c_i(\cP)$ corresponds to a feasible subtree $\bT'$ of $\bT$ with the cost $f_i(\bT')$ such that $f_i(\bT')=c_i(\cP)$ and vice versa.
\label{obs:sp-graph:subtree=stpath}
\end{observation}

\subsection{LP Formulation}
\label{sec:sp-graph:lp}

Given any series-parallel graph $G:=(V,E)$, we first employ the standard doubling technique to enhance the linear program.
Given a guess of the optimal objective value $\gs$, we discard those edges $e$ such that there exists an agent $i\in[k]$ with $c_i(e)>\gs$.
Clearly, these discarded edges cannot belong to the optimal solution. 
Then, we run the series-parallel graph recognition algorithm~\cite{DBLP:journals/siamcomp/ValdesTL82} to construct a decomposition tree $\bT:=(\bV,\bE)$ of the series-parallel graph.
See \cref{alg:sp-graph} for the complete description.

The linear program is shown in~\eqref{sp:tree-LP}. 
For an internal node $\bv\in \bV$, use $\kid(\bv)$ and $\Lambda(\bv)$ to denote its children and descendants in the tree respectively. 
Let $P(\bT)$ and $S(\bT)$ be the set of parallel and series nodes in $\bT$.
For each node $\bv$, $\xv$ is a relaxed indicator variable denoting whether $\bv$ is selected or not. 
The three constraints~\eqref{sp:tree-LPC:root},~\eqref{sp:tree-LPC:parallel} and~\eqref{sp:tree-LPC:series} correspond to the three conditions stated in~\cref{def:sp:feasible-subtree} respectively, in order to ensure that the solution is a feasible subtree. 
The first constraint type~\eqref{sp:tree-LPC:cost} is a bit subtle, and it is the key that allows us to surpass the pessimistic $\Omega(\max\{k,\sqrt{n}\})$ integrality gap. In these constraints, $\sum_{\bu\in \Lambda(\bv)}\xu \cdot f_i(\bu)$ denotes the cost of the selected subtree rooted at $\bv$ with respect to agent $i$. 
Thus, when $\bv=\br$, the constraint implies that for any agent, the total cost of all the selected nodes must be at most $x_{\br}\cdot\opt=\opt$. 
For the cases that $\bv\neq \br$, these constraints do not affect the feasible region of integer solutions since $x_{\bv}$ is either $1$ or $0$, but they can reduce the fractional solution's feasible region dramatically by restricting the contribution of each subtree $\Lambda(\bv)$. 
A more detailed discussion is given in~\cref{sec:tree-lp:discussion}.

\begin{align}
    &&  & \tag{\text{Tree-LP}} \label{sp:tree-LP}\\
    &&\sum_{\bu\in \Lambda(\bv)}\xu \cdot f_i(\bu) &\leq \xv \cdot \gs, &\forall i\in [k], \forall \bv\in \bV\label{sp:tree-LPC:cost} \\
    &&\xr &=1 , & \label{sp:tree-LPC:root}\\
    &&\sum_{\bu\in\kid(\bv)}\xu &= \xv, &\forall \bv\in P(\bT) \label{sp:tree-LPC:parallel}\\
    &&\xu &= \xv, & \forall \bv\in S(\bT), \bu\in\kid(\bv) \label{sp:tree-LPC:series}\\
    &&\xv &\geq 0, &\forall \bv\in \bV 
\end{align}

\subsection{Algorithms}
\label{sec:sp-graph:alg}

This section formally describes the complete algorithm (\cref{alg:sp-graph}) for series-parallel graphs.
The main algorithm mainly consists of two steps: the doubling step (lines \ref{lines:sp:binary-search-1}-\ref{lines:sp:binary-search-2} of \cref{alg:sp-graph}) and the rounding algorithm (\cref{alg:rounding}).
After finishing the doubling step, we obtain a fractional solution $x^*$ with the value of $\gs$ that is close to the optimal solution $\opt$ (\cref{obs:sp-graph:guess}).
And then, we shall employ the same rounding algorithm in~\cite{DBLP:journals/corr/abs-2302-11475} to obtain a feasible subtree based on $x^*$.
This dependent randomized rounding algorithm selects nodes level by level, starting from the top of $\bT$ and proceeding downwards. 
For parallel nodes, the algorithm selects one of its child nodes with a probability determined by the optimal fractional solution $x^*$. 
For series nodes, the algorithm selects all of its child nodes with a probability of $1$, ensuring that the resulting subtree is always feasible. 
A formal description of the algorithm can be found in \cref{alg:rounding}.

\begin{algorithm}[htb] 
\caption{The Complete Algorithm for Series-Parallel Graphs}
\label{alg:sp-graph}
\begin{algorithmic}[1]
\Require A series-parallel graph $G:=(V,E)$ with $k$ cost functions $c_i:2^{E}\to \R_{\geq 0},i\in[k]$.
\Ensure An $s \tto t$ path $\cP\subseteq E$.
\State $\flag \leftarrow \true$; $\gs \leftarrow \max_{i\in[k]}\sum_{e\in E}c_i(e)$.
\While{$\flag = \true$}
\label{lines:sp:binary-search-1}
\State $E'\leftarrow\set{e\in E \mid \exists i\in[k]\text{ s.t. }c_i(e)>\gs}$.
\State $E\leftarrow E\setminus E'$; $G\leftarrow (V,E')$.
\State Compute the tree decomposition $\bT:=(\bV,\bE)$ of $G$ by~\cite{DBLP:journals/siamcomp/ValdesTL82}.
\State Solve the linear program \eqref{sp:tree-LP}.
\If{\eqref{sp:tree-LP} has a feasible solution}
\State Let $x^*$ be a feasible solution to \eqref{sp:tree-LP}.
\State $\gs \leftarrow \frac{\gs}{2}$.
\Else
\State $\flag \leftarrow \false$.
\EndIf
\EndWhile
\label{lines:sp:binary-search-2}
\State Run \cref{alg:rounding} with the optimal solution $x^*$ to obtain a feasible subtree $\bT'$ of $\bT$.
\State Convert $\bT'$ into an $s \tto t$ path $\cP$.
\State \Return $\cP$.
\end{algorithmic}
\end{algorithm}

\begin{algorithm}[htb] 
\caption{Dependent Randomized Rounding}
\label{alg:rounding}
\begin{algorithmic}[1]
\Require A tree structure $\bT(\bV,\bE)$ rooted at $\br$; a fractional solution $x^*\in[0,1]^{|\bV|}$.
\Ensure A feasible subtree $\bT'$.
\State Initially, set $\bT' \leftarrow \emptyset$ and a vertex queue $\cQ \leftarrow \{\br\}$.
\While{$\cQ \neq \emptyset $}
\State Use $\bv$ to represent the front element of $\cQ$.
\State $\bT' \leftarrow \bT' \cup \{\bv\}$, $\cQ \leftarrow \cQ \setminus \{\bv\}$.
\If{$\bv$ is a parallel node}
\State Pick exactly one node $\bu\in\kid(\bv)$ randomly such that $\bu$ is chosen with probability $\frac{\xu}{\xv}$.
\State $\cQ \leftarrow \cQ \cup \{\bu\}$.
\EndIf
\If{$\bv$ is a series node}
\For{each $\bu\in\kid(\bv)$}
\State $\cQ \leftarrow \cQ \cup \{\bu\}$.
\EndFor
\EndIf
\EndWhile
\State \Return $\bT'$.
\end{algorithmic}
\end{algorithm}

\subsection{Analysis}

This section analyzes the performance of our algorithm. We start by describing a simple observation (\cref{obs:sp-graph:guess}). 
Let $\bT'$ be the subtree returned by \cref{alg:rounding}.
Recall that $H$ is the height of $\bT$.

\begin{observation}
Let $\gs^*$ be the guessing value at the beginning of the last round of the while-loop (lines \ref{lines:sp:binary-search-1}-\ref{lines:sp:binary-search-2} of \cref{alg:sp-graph}). 
Then, we have $\gs^* \leq 2\cdot \opt$\footnote{One can get a more accurate lower bound of the optimal solution (e.g., $\gs^*\leq (1+\epsilon) \cdot \opt$ for any $\epsilon >0$) by the standard binary search technique.}.
\label{obs:sp-graph:guess}
\end{observation}

\begin{proof}
Let $\lp$ be the minimum value of $\gs$ such that \eqref{sp:tree-LP} can have a feasible solution.
Clearly, $\lp$ is a lower bound of the optimal solution, i.e., $\lp \leq \opt$.
Now we consider the last round of the while-loop, which implies that the current value of $\gs^*$ is still large enough to make \eqref{sp:tree-LP} admit a feasible solution.
Since the current while-loop is the last round, we know that if we cut edges according to $\frac{\gs^*}{2}$, \eqref{sp:tree-LP} will not have a feasible solution.
Thus, we have $\frac{\gs^*}{2}\leq \lp \leq \opt$.
\end{proof}

To show that the approximation ratio is $O(H\log k)$ with a constant probability, a natural step is to first bound the expectation of our solution. We first state some intuition. According to the description of the rounding scheme, it is easy to see that for each agent $i$, 
\[\E[f_i(\bT')]=\sum_{e\in E}c_i(e)\Pr[e\in \bT']=\gs^*\].
Then by Markov inequality, we have for each agent $i$,
\[ \Pr[f_i(\bT') \geq H\log k \cdot \gs^* ] \leq \frac{1}{H\log k}.\]
However, the above inequality is not sufficient because proving \cref{thm:sp-graph} needs to show that $\Pr[\forall i\in [k], f_i(\bT') \geq H\log k \cdot \gs^*]$ is at most $\frac{1}{k}+\frac{1}{kH}$. To address this issue, we need to employ an analysis technique called \emph{Moment Method}, which is widely used in the literature~\cite{DBLP:journals/corr/abs-2302-11475,DBLP:conf/stoc/0001LL19}. 
More formally, we aim to show the following key lemma (\cref{lem:s-t:concentration}); a similar proof can also be found in~\cite{DBLP:journals/corr/abs-2302-11475}.

\begin{lemma}
For any agent $i\in [k]$, we have $\E[\exp\left({\ln (1+\frac{1}{2H})\cdot f_i(\bT')}\right)] \leq 1+\frac{1}{H}$.
\label{lem:s-t:concentration}
\end{lemma}

\begin{proof}
    We prove the theorem inductively. First, consider the case that $H=1$, i.e., the decomposition tree $\bT$ only contains a root $\br$. Since $x_{\br}=1$, there is no randomness in selecting $\bT'$. Thus, we have for any $z\geq 1$,
    \[\E\left[z^{\frac{f_i(\bT')}{\gs^*}}\right] = z^{\frac{x_{\br} \cdot f_i(\br)}{\gs^*}} \leq z,\]
    where the last inequality uses constraint~\eqref{sp:tree-LPC:cost} in \eqref{sp:tree-LP}.

    To streamline the analysis, we continue considering the case that $H=2$. We further distinguish two subcases: (\rom{1}) root $\br$ is a parallel node; (\rom{2}) root $\br$ is a series node. For the first subcase,~\cref{alg:rounding} selects exactly one child $\bv\in \kid(\br)$ with probability $x_{\bv}/x_{\br} = x_{\bv}$. According to the law of total expectation and only leaves in $\bT$ have non-zero costs, we have
    \begin{equation*}
        \begin{aligned}
           \E\left[z^{\frac{f_i(\bT')}{\gs^*}}\right] = \sum_{\bv\in \kid(\br)} \Pr[\bv \in \bT'] \cdot \E\left[z^{\frac{f_i( \Lambda'(\bv) )}{\gs^*}} \middle| \bv \in \bT' \right] ,
        \end{aligned}
    \end{equation*}
    where $\Lambda'(\bv) := \Lambda(\bv) \cap \bT'$. Observing that once conditioned on $\bv \in \bT'$, the conclusion for the $H=1$ case can be used to bound the expectation, we have
    \begin{align*}
        \E\left[z^{\frac{f_i(\bT')}{\gs^*}}\right] & = \sum_{\bv\in \kid(\br)} \Pr[\bv \in \bT'] \cdot \E\left[z^{\frac{f_i( \Lambda'(\bv) )}{\gs^*}} \middle| \bv \in \bT' \right] , \\
        & = \sum_{\bv\in \kid(\br)} x_{\bv} \cdot z^{\frac{f_i( \bv )}{\gs^*}}  \\
         & \leq \sum_{\bv\in \kid(\br)} x_{\bv} \cdot \left(1+(z-1)\cdot \frac{f_i( \bv )}{\gs^*}\right) \tag{Constraint~\eqref{sp:tree-LPC:cost} and $z^r \leq 1+r(z-1) \forall z > 0, r\in [0,1]$ } \\
          & = \left(\sum_{\bv\in \kid(\br)} x_{\bv}\right) + (z-1)\cdot  \frac{\sum_{\bv\in \kid(\br)} x_{\bv}\cdot f_i(\bv)}{\gs^*} \\
          & = 1 + (z-1)\cdot \frac{\sum_{\bv\in \Lambda(\br)} x_{\bv}\cdot f_i(\bv)}{\gs^*} \tag{Constraint~\eqref{sp:tree-LPC:parallel}}\\ 
          & \leq e^{(z-1)\cdot \frac{\sum_{\bv\in \Lambda(\br)} x_{\bv}\cdot f_i(\bv)}{\gs^*} } \tag{$1+x\leq e^x$}.
    \end{align*}
           
    For the second subcase, according to Constraint~\eqref{sp:tree-LPC:series}, we have $x_{\bv}=x_{\br}=1$ for each $x_{\bv}\in \kid(\br)$, and therefore, 
    \begin{align*}
        \E\left[z^{\frac{f_i(\bT')}{\gs^*}}\right]  = \E\left[z^{\frac{\sum_{\bv \in \kid(\br) f_i(\Lambda'(\bv))}  }{\gs^*}}\right] \leq z^{ \frac{\sum_{\bv\in \Lambda(\br)} x_{\bv}\cdot f_i(\bv)}{\gs^*} }.
    \end{align*}

    Since $z\leq e^{z-1}$, combing the two subcases, we have that when $H=2$,
    \[\E\left[z^{\frac{f_i(\bT')}{\gs^*}}\right] \leq (e^{(z-1)})^{ \frac{\sum_{\bv\in \Lambda(\br)} x_{\bv}\cdot f_i(\bv)}{\gs^*} }.  \]

    The above inequality shows that as the height increases by $1$, the upper bound of the expectation grows exponentially. Furthermore, when the height increases from $1$ to $H$, we can obtain a sequence $z_1=z,z_2,\ldots,z_H$, where $z_{h} = e^{z_{h-1}-1}$ for each $h>1$, and have
    \[\E\left[z^{\frac{f_i(\bT')}{\gs^*}}\right] \leq z_h^{ \frac{\sum_{\bv\in \Lambda(\br)} x_{\bv}\cdot f_i(\bv)}{\gs^*} }, \]
    for any $\bT$ with height $h$ by almost the same analysis as above. Due to Constraint~\eqref{sp:tree-LPC:cost}, we have $\sum_{\bv\in \Lambda(\br)} x_{\bv}\cdot f_i(\bv) \leq \gs^*$, and thus, $\E\left[z^{\frac{f_i(\bT')}{\gs^*}}\right]$ is at most $z_h$.
    
    Finally, to obtain the claimed upper bound, we set $z=1+\frac{1}{2H}$ such that $z_h\leq 1+\frac{1}{H}$, and complete the proof.
\end{proof}

\begin{lemma}
Consider an arbitrary agent $i$, \cref{alg:rounding} returns a feasible subtree $\bT'$ such that $f_i(\bT') \leq 4H\cdot \log k \cdot \gs$ with high probability, where $H$ is the height of the decomposition tree of the series-parallel graph and $\gs$ is a guess of the optimal objective value such that the corresponding \eqref{sp:tree-LP} admits a feasible solution.
\label{lem:sp-graph:key}
\end{lemma}

\begin{proof}
It is easy to see that $\bT'$ must be a feasible subtree.
We have the following inequalities:
\begin{align*}
&\E\left[\left(1+\frac{1}{2H}\right)^{f_i(\bT')/\gs}\right] \leq 1+\frac{1}{H} \tag*{[By \cref{lem:s-t:concentration}]} \\
\Rightarrow & \Pr\left[ \left(1+\frac{1}{2H}\right)^{f_i(\bT')/\gs} \geq k^2 \right] \leq \frac{1}{k^2}+\frac{1}{k^2 H} \tag*{[By Markov bound]} \\
\Rightarrow & \Pr\left[ f_i(\bT') \cdot \log(1+\frac{1}{2H}) \geq 2\log k \cdot \gs \right] \leq \frac{1}{k^2}+\frac{1}{k^2 H} \\
\Rightarrow & \Pr\left[ f_i(\bT') \geq 8H\log k \cdot \gs \right] \leq \frac{1}{k^2}+\frac{1}{k^2 H} \tag*{[By $\log(1+\frac{1}{2H}) \geq \frac{1}{4H}$]} 
\end{align*}
\end{proof}

\begin{proofof}{\cref{thm:sp-graph}}
After finishing the doubling step, we can get a guess $\gs^*$ such that $\gs^*$ can make \eqref{sp:tree-LP} admit a feasible solution.
By \cref{lem:sp-graph:key}, we know that there is a randomized algorithm that outputs a feasible subtree such that, for each $i\in[k]$:
$$
\Pr\left[ f_i(\bT') \cdot \geq 8H\log k \cdot \gs^* \right] \leq \frac{1}{k^2}+\frac{1}{k^2 H}.
$$
By union bound, we have:
$$
\Pr\left[ \max_{i\in[k]}f_i(\bT') \cdot \leq 8H\log k \cdot \gs^* \right] \geq 1- \left(  \frac{1}{k}+\frac{1}{k H} \right).
$$
By \cref{obs:sp-graph:subtree=stpath}, we know that $\bT'$ corresponds to an $s \tto t$ path $\cP$ such that $c_i(\cP)=f_i(\bT')$ for all $i\in[k]$.
Thus, we have:
$$
\Pr\left[ \max_{i\in[k]}c_i(\cP)=\max_{i\in[k]}f_i(\bT') \cdot \leq 8H\log k \cdot \gs^* \right] \geq 1- \left(  \frac{1}{k}+\frac{1}{k H} \right).
$$
By \cref{obs:sp-graph:guess}, we have $\gs^* \leq 2\cdot \opt$ and thus we have:
$$
\Pr\left[ \max_{i\in[k]}c_i(\cP)\cdot \leq 16 \cdot H \cdot \log k \cdot \opt \right] \geq 1- \left(  \frac{1}{k}+\frac{1}{k H} \right).
$$
\end{proofof}

\section{General Graphs}
\label{sec:general-graphs}

We remark that the running time of \cref{alg:treewidth} is $O(n^n)$ for general graphs since the treewidth of the general graph may be as large as $n$.
Thus, we need to seek other methods to obtain a more efficient algorithm.
Due to the substantial integrality gap observed in the standard linear programming relaxation stated in \cref{sec:flow-lp-gap}, we are compelled to explore alternative formulations that can effectively eliminate the gap of $\Omega(\max\{k,\sqrt{n}\})$.
We shall follow the same algorithmic framework stated in \cref{sec:sp} and show the following result (\cref{thm:general:graph}).

\begin{theorem}
Given any directed graph $G$, there is an algorithm that returns a $O(\log n \log k)$-approximate solution in $\poly(n)\cdot n^{O(\log n)}$ time with probability at least $1-(\frac{1}{k}+\frac{1}{k \log n})$ for robust  $s\tto t$ path, where $n$ is the number of vertices and $k$ is the number of agents.
\label{thm:general:graph}
\end{theorem}

In the following, we first briefly recall the algorithmic framework used in \cref{sec:sp}, and then we provide some intuition and formal description of the tree construction in \cref{sec:supertree}.
We then give the LP formulation for general graphs in \cref{sec:general-graph:lp}.
Finally, we present the complete description of our algorithm and analysis in \cref{sec:general-graph:alg}.

\paragraph{Algorithmic Framework.}
Same as \cref{sec:sp}, our algorithm (\cref{alg:general-graph}) is mainly divided into two parts: tree construction and rounding algorithm.
For general graphs, we no longer have a simple and natural tree structure like the series-parallel graph.
To use the same rounding algorithm (\cref{alg:rounding}), we aim to construct a decision-tree-type metatree structure that maps every $s \tto t$ path in the given directed graph to a corresponding subtree in the metatree. 
Although the size of the metatree is not polynomial with respect to $n$, we are able to establish that a quasi-polynomially sized metatree is adequate to enclose all $s \tto t$ paths of the original directed graph. 
This explains why our algorithm runs efficiently in quasi-polynomial time. 
The metatree constructed in this way is extensive, and not all subtrees correspond to an $s \tto t$ path of the original graph. 
In the second part, we present a linear program formulation derived from the metatree that guarantees that the output subtree can be converted into an $s \tto t$ path. 
Finally, we show that \cref{alg:rounding} achieves an approximation factor of $O(\log n \log k)$ using a similar proof as the \cref{thm:sp-graph}.

\subsection{Metatree Construction}
\label{sec:supertree}

In this section, we outline the steps involved in constructing the metatree $\bT:=(\bV,\bE)$. 
Our goal is to ensure that (\rom{1}) every $s \tto t$ path in the directed graph $G:=(\abs{V}=n,\abs{E}=m)$ corresponds to a subtree, and (\rom{2}) the size of $\bT$ is $n^{O(\log n)}$, i.e., $\abs{\bV}=n^{O(\log n)}$.
To achieve this, we select an index of agent $i\in[k]$ and define how to map the cost function $c_i$ from $G$ to $\bT$. 
For ease of notation, we omit the agent index from the cost function and use $c$ to represent $c_i$ in this section.

\paragraph{Metatree Construction Intuition.}
The basic idea of the metatree construction is to iteratively guess the possible middle vertex of an $s\tto t$ path. 
There are at most $n$ possibilities for this middle vertex. 
Once we determine the middle vertex of the $s\tto t$ path, say it is $a$, the whole path can be partitioned into two subpaths---path $s\tto a$ and path $a\tto t$.
We then recur on the obtained subpaths till level $O(\log n)$; the sufficiency of $O(\log n)$'s levels will be clear later.
This process gives us a natural tree structure $\bT$ with $O(\log n)$ depth. 
We define two types of nodes in $\bT$. 
The first node type is referred to as {\em splitting node}. 
Each splitting node corresponds to a (sub-)path. 
It has $n$ children, where each child represents a choice of the (sub-)path's middle vertex. 
The algorithm needs to ensure that only one of these children can be selected. 
The second node type is called {\em merging node}. 
A merging node has to be a child of a splitting node in $\bT$ and represents a scheme for selecting a middle vertex. 
Further, a merging node has two splitting nodes as its children, corresponding to the two obtained subpaths by this scheme. 
We can view such a node as being used to merge its two children (subpaths). 
The algorithm needs to ensure that both children are selected simultaneously. 
A similar decision-tree-type structure construction method can be found in~\cite{DBLP:journals/corr/abs-2302-11475,DBLP:conf/stoc/0001LL19}.

As one may observe, a splitting (resp. merging) node in our metatree plays the same role as a parallel (resp. series) node in the decomposition tree of the series-parallel graph.
Thus, the LP for the general graph is similar to \eqref{sp:tree-LP} and we can still use the same rounding algorithm.
Consider an $s\tto t$ path $\cP$ of length $n$.
If we write $\cP$ in the form of a binary tree by guessing the middle vertex of each subpath, it will have at most $\ceil{\log n}$ levels.
Thus, we can let the recursive tree $\bT$ terminate at level $O(\log n)$, and therefore, its size is quasi-polynomial $O(n^{\log n})$ since each node has at most $n$ children. 
From \cref{sec:sp}, we know that the rounding algorithm is able to find a $O(H \log k)$-approximate solution where $H$ is the height of the tree.
This also provides the intuition for the approximation ratio $O(\log n \log k)$ since the height of $\bT$ is $O(\log n)$ (specifically, $H=2\ceil{\log n}+1$).

\paragraph{Metatree Construction.}
Formally, each node in $\bV$ is either a splitting node or a merging node.
In addition, we assign a label to each node in $\bV$ which is used to characterize its meaning and associate a cost of the node.
If a node $\bv\in \bV$ is a splitting node, its label $\lbl(\bv)$ is a 2-tuple: $u\tto w$ that indicates the path from $u$ to $w$ of the original graph $G$.
If a node $\bv\in \bV$ is a merging node, its label $\lbl(\bv)$ is a 3-tuple: $u\tto q \tto w$ that indicates two subpaths of the path from $u$ to $w$, i.e., the path from $u$ to $q$ and the path from $q$ to $w$.
Since we only care about the $s \tto t$ path, either $u$ is the source $s$ or $w$ is the sink $t$.
Note that each splitting node has exactly $n$ children, each of which is a merging node, and each merging node has exactly two children, each of which is a splitting node. 
Let $\br$ be the root of $\bT$, which is a splitting node with label $\lbl(\br)=s\tto t$. 
Then, we feed $\br$ to \cref{alg:supertree} to obtain the desired metatree $\bT$.

\begin{algorithm}[htb] 
\caption{$\texttt{tree-construction}(\bv)$}
\label{alg:supertree}
\begin{algorithmic}[1]
\Require A directed acyclic graph $G=(V,E)$; The root $\br$ of $\bT$.
\Ensure A metatree $\bT$.
\If{the number of levels of $\bT$ is strictly larger than $2\ceil{\log n}+1$}
\State \Return $\bT$.
\Else
\If{$\bv$ is a splitting node}
\For{each $w\in V$}
\State creates a merging node $\bp$ as $\bv$'s child.
\State $\lbl(\bp) \leftarrow a\tto w \tto b$. 
\Comment{Suppose $\lbl(\bv)=a\tto b$.}
\State {\bf call} $\texttt{tree-construction}(\bp)$.
\EndFor
\EndIf
\If{$\bv$ is a merging node}
\State creates a splitting node $\bp$ as $\bv$'s left child and another splitting node $\bp'$ as its right child.
\State $\lbl(\bp) \leftarrow a\tto c$ and $\lbl(\bp')\leftarrow c \tto b$.
\Comment{Suppose $\lbl(\bv)=a\tto c \tto b$.}
\State {\bf call} $\texttt{tree-construction}(\bp)$.
\State {\bf call} $\texttt{tree-construction}(\bp')$. 
\EndIf
\EndIf
\end{algorithmic}
\end{algorithm}

We now define a {\em feasible subtree} for $\bT$.
Same to the series-parallel graph, if a subtree $\bT'$ of $\bT$ corresponds to an $s\tto t$ path, then $\bT'$ is called a {\em feasible subtree}.
However, we cannot use the same definition as \cref{def:sp:feasible-subtree} to characterize the feasible subtree for general graphs.
This is because not all leaf nodes in the constructed tree $\bT$ correspond to edges in $G$.
We refer to a subtree that satisfies three conditions in \cref{def:sp:feasible-subtree} as a {\em consistent subtree}. 
To ensure that the subtree can be translated to an $s\tto t$ path, one more condition is needed.

\begin{definition}[Feasible Subtree for General Graphs]
A subtree $\bT' \subseteq \bT$ is called a feasible subtree if and only if (\rom{1}) $\bT'$ is a consistent subtree; (\rom{2}) each leaf node corresponds to either an edge or a single vertex in $G$.
\label{def:feasible-subtree}
\end{definition}

\begin{figure}[htb]
    \centering
    \includegraphics[width=15cm]{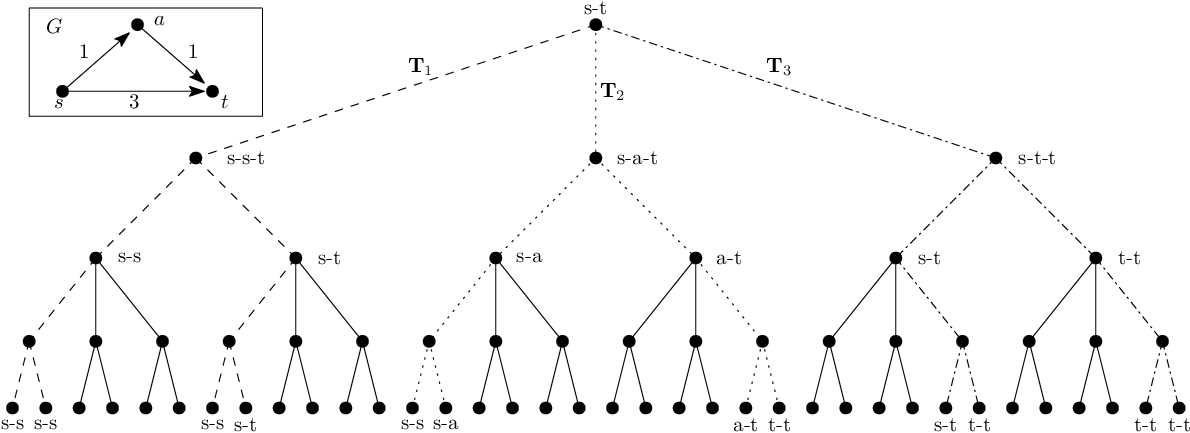}
    \caption{An example for the metatree $\bT$ construction. The given directed graph $G$ is shown in the up-left corner. Since there are three vertices in $G$, $\bT$ will consist of five levels. The dashed subtree $\bT_1$ corresponds to the path $s\to t$ of $G$. The dotted subtree $\bT_2$ represents the path $s\to a \to t$ of $G$. The dash-dotted subtree $\bT_3$ also corresponds to the path $s\to t$. By the definition of the cost function $f$, it is not hard to verify that the cost of each subtree is equal to the cost of its corresponding path.} 
    \label{fig:supertree}
\end{figure}

An example can be found in \cref{fig:supertree}.
As one might be concerned, using a different definition of the feasible subtree may require a different LP formulation for general graphs, since a natural adaptation of \eqref{sp:tree-LP} can only find a consistent subtree. We show that this issue can be fixed easily by directly disabling the infeasible leaf nodes in the linear program. 
See \cref{sec:general-graph:lp} for more details of the LP formulation.

Now, it remains to assign a cost to each node in $\bV$.
Recall that we have a cost function $c:2^E\to\R_{\geq 0}$ defined over the edges of the original directed graph.
We will construct a cost function $f:2^{\bV}\to\R_{\geq 0}$ according to the function $c$.
Note that $\bT$ has exactly $2\ceil{\log n}+1$ levels by the definition of \cref{alg:supertree}.
Thus, all nodes in $\leaf(\bT)$ must be splitting nodes since the first level is the root which is a splitting node and $2\ceil{\log n}+1$ is an odd number.
Let $\leaf(\bT) \subseteq \bV$ be the set of leaf nodes of $\bT$.
We consider an arbitrary node $\bv$ in $\leaf(\bT)$ and assume that its label $\lbl(\bv)$ is $a\tto b$.
Now, we define the cost function $f$ as follows: (\rom{1}) $f(\bv)=0$ if $v\notin\leaf(\bT)$; (\rom{2}) $f(\bv)=0$ if $\bv\in\leaf(\bT)$ and $a,b$ are the same node; (\rom{3}) $f(\bv)=+\infty$\footnote{In fact, $f(\bv)$ can be an arbitrary value in this case, since the LP variable of such a vertex is always $0$; see \eqref{tree-LPC:infeasible} of \eqref{tree-LP}.} if $\bv\in\leaf(\bT)$ and $(a,b)\notin E$; (\rom{4}) $f(\bv)=c((a,b))$ if $\bv\in\leaf(\bT)$ and edge $(a,b)\in E$.
By defining this specific cost function, we show in \cref{lem:s-t:feasibility} that the cost of an arbitrary $s \tto t$ path of $G$ is precisely the cost of its corresponding subtree of $\bT$.
An example can be found in \cref{fig:supertree}.

\begin{lemma}
Consider an $s \tto t$ path $\cP\subseteq E$ of the given directed graph $G=(V,E)$ such that $\cP$ does not go through the same vertex more than once.
Then, there exists a feasible subtree $\bT'\subseteq \bV$ of $\bT=(\bV,\bE)$ such that $c(\cP)=f(\bT')$.
\label{lem:s-t:feasibility}
\end{lemma}

\begin{proof}
We consider the $s \tto t$ path $\cP:=(s\to 1 \to \cdots \to \ell \to t)$ of $G$.
Note that the length of $\cP$ is at most $n$ since $\cP$ does not go through the same vertex twice.
Note that $\cP$ may correspond to many feasible subtrees of $\bT$.
To show \cref{lem:s-t:feasibility}, it is sufficient to construct a feasible subtree with the same cost.
To this end, we shall follow the basic construction idea for $\bT$ to find a feasible subtree for $\cP$, i.e., guessing the middle vertex of $\cP$.
We start from the root of $\bT$, and choose nodes level by level from the top to the bottom.
If the current node $\bv$ is a merging node, then choose all its child nodes.
If the current node $\bv$ is a splitting node (suppose that $\lbl(\bv)=s\tto a$), then we always choose $\bv$'s children node with label $s \tto d \tto a$ where vertex $d$ is the middle vertex in the path from $s$ to $a$.
The case where $\lbl(\bv)=a\tto t$ is the same, i.e., we always choose the middle vertex in the path from $a$ to $t$.
Note that it is possible that $s\tto a$ (or $a \tto t$) is an edge in $G$.
In this case, we just pick the node with the label of $s\tto s \tto a$ (or $a \tto a \tto t$).
We repeatedly choose the nodes according to the above rule until we reach the leaf node of $\bT$.
Let $\bT'$ be the constructed subtree.
Consider an arbitrary leaf node $\bv$ of $\bT'$, i.e., $\bv\in\leaf(\bT)$.
Suppose that $\lbl(\bv)=f\tto h$.
It is easy to see that either $(f,h)$ is an edge in $\cP$ or $f,h$ are the same vertex.
Then, by the definition of the cost function in $\bT$, we know that $c(\cP)=f(\bT')$.
\end{proof}

\subsection{LP Formulation}
\label{sec:general-graph:lp}

Given a node $\bv$ in $\bV$, let $\kid(\bv)\subseteq \bV$ be the set of child nodes of $v$.
Let $S(\bT)$ and $M(\bT)$ be the set of splitting nodes and merging nodes in $\bV$.
Note that $S(\bT) \cap M(\bT) = \emptyset$ and $S(\bT)\cup M(\bT)=\bV$.
For each node $\bv$ in $\bV$, we create a binary variable $\xv\in\set{0,1}$, where $\xv=1$ indicates that $\bv$ is selected and $\xv=0$ indicates that it is not. 
We relax the integral constraints of $\xv$ by allowing $\xv\in[0,1]$.
We then follow the basic idea of \eqref{En-Flow-LP} which combines the binary search and truncated instance.
Given a guess of the optimal objective value $\gs$, we discard those edges $e$ such that there exists an agent $i\in[k]$ with $c_i(e)> \gs$.
The formal description can be found in \cref{alg:general-graph}.
Then, we run \cref{alg:supertree} to construct a metatree structure for this truncated instance.
We solve the following LP relaxation (\ref{tree-LP}) to check whether there exists a feasible solution.
Given a node $\bv\in \bV$, let $\Lambda(\bv)$ be the set of nodes included in the subtree that is rooted by $\bv$.
Note that $\Lambda(\bv)$ also includes $\bv$ itself.
Let $\Phi(\bT)\subseteq\leaf(\bT)$ be the set of leaf nodes that do not correspond to an edge in $E$, i.e., for each node $\bv\in\leaf(\bT)$ with $\lbl(\bv)=a\tto b$, $\bv\in \Phi(\bT)$ if and only if $(a,b)\notin E$ and $a\ne b$.

\begin{align}
     &&  & \tag{\text{GG-Tree-LP}} \label{tree-LP}\\
    &&\sum_{\bu\in \Lambda(\bv)}\xu \cdot f_i(\bu) &\leq \xv \cdot \gs, &\forall i\in [k], \forall \bv\in \bV\label{tree-LPC:cost} \\
    &&\xr &=1 , & \label{tree-LPC:root}\\
    &&\sum_{\bu\in\kid(\bv)}\xu &= \xv, &\forall \bv\in S(\bT) \label{tree-LPC:decision}\\
    &&\xu &= \xv, & \forall \bv\in M(\bT), \bu\in\kid(\bv) \label{tree-LPC:merge}\\
    &&\xv &= 0, & \forall \bv\in\Phi(\bT) \label{tree-LPC:infeasible}\\
    &&\xv &\geq 0, &\forall \bv\in \bV 
\end{align}

\eqref{tree-LPC:cost} are the cost constraints.
\eqref{tree-LPC:root} ensure that the root of $\bT$ must be included in the solution.
\eqref{tree-LPC:decision} and \eqref{tree-LPC:merge} are the constraints for the splitting and merging nodes, which guarantee that the selected subtree is a consistent subtree.
Based on \eqref{tree-LPC:decision} and \eqref{tree-LPC:merge}, \eqref{tree-LPC:infeasible} ensures that the selected subtree is a feasible subtree.

\subsection{Algorithm \& Analysis}
\label{sec:general-graph:alg}

In this section, we formally describe the complete algorithm (\cref{alg:general-graph}) for general graphs.
Same as \cref{alg:sp-graph}, \cref{alg:general-graph} mainly consists of two parts: doubling and rounding.
The only difference between \cref{alg:sp-graph} and \cref{alg:general-graph} is that we need to employ another method to construct the tree structure (i.e., \cref{alg:supertree}).
Note that the splitting and merging node in the current section correspond to the parallel and series node in \cref{sec:sp}. 

\begin{algorithm}[htb] 
\caption{The Complete Algorithm for General Graphs}
\label{alg:general-graph}
\begin{algorithmic}[1]
\Require A general $G:=(V,E)$ with $k$ cost functions $c_i:2^{E}\to \R_{\geq 0},i\in[k]$.
\Ensure An $s \tto t$ path $\cP\subseteq E$.
\State $\flag \leftarrow \true$; $\gs \leftarrow \max_{i\in[k]}\sum_{e\in E}c_i(e)$.
\While{$\flag = \true$}
\label{lines:gp:binary-search-1}
\State $E'\leftarrow\set{e\in E \mid \exists i\in[k]\text{ s.t. }c_i(e)>\gs}$.
\State $E\leftarrow E\setminus E'$; $G\leftarrow (V,E')$.
\State Construct the metatree $\bT:=(\bV,\bE)$ of $G$ by \cref{alg:supertree}.
\State Solve the linear program \eqref{tree-LP}.
\If{\eqref{tree-LP} has a feasible solution}
\State Let $x^*$ be a feasible solution to \eqref{tree-LP}.
\State $\flag \leftarrow \true$; $\gs \leftarrow \frac{\gs}{2}$.
\Else
\State $\flag \leftarrow \false$.
\EndIf
\EndWhile
\label{lines:gp:binary-search-2}
\State Run \cref{alg:rounding} with the optimal solution $x^*$ to obtain a feasible subtree $\bT'$ of $\bT$.
\State Convert $\bT'$ into an $s \tto t$ path $\cP$.
\State \Return $\cP$.
\end{algorithmic}
\end{algorithm}

Let $\bT'\subseteq \bV$ be the subtree returned by \cref{alg:rounding}.
Then, the subtree $\bT'$ has two simple but useful properties (\cref{lem:s-t:alg-prop}).
Let $H:=2\cdot \ceil{\log n}+1$ be the height of $\bT$.

\begin{observation}
Let $\gs^*$ be the guessing value at the beginning of the last round of the while-loop (lines \ref{lines:gp:binary-search-1}-\ref{lines:gp:binary-search-2} of \cref{alg:general-graph}). 
Then, we have $\gs^* \leq 2\cdot \opt$.
\label{obs:gp:guess}
\end{observation}

\begin{lemma}
Let $\bT'$ be the subtree returned by \cref{alg:rounding}.
Then, $\bT'$ must be a feasible subtree.
\label{lem:s-t:alg-prop}
\end{lemma}

\begin{proof}
It is not hard to see that the returned subtree must be a consistent subtree since the algorithm selects the types of nodes level by level.
And \eqref{tree-LPC:infeasible} in \eqref{tree-LP} ensures that the selected consistent subtree must not contain any nodes that do not correspond to an edge or vertex.
By \cref{lem:s-t:feasibility}, we know that any non-redundant $s\tto t$ path (the path goes through one vertex more once) of the original graph corresponds to a feasible subtree of $\bT$ with the same cost.
Thus, there must exist a feasible subtree of $\bT$ with a bounded cost.
This proves the property of the lemma since the input of \cref{alg:rounding} is a nearly optimal solution.
\end{proof}

\begin{observation}
For any agent $i\in [k]$, we have \[\E\left[\left(1+\frac{1}{2H}\right)^{f_i(\bT')/\gs}\right] \leq 1+\frac{1}{H},\] 
where $H$ is the height of the decomposition tree of the series-parallel graph and $\gs$ is a guess of the optimal objective value such that the corresponding \eqref{tree-LP} admits a feasible solution.
\label{obs:s-t:concentration}
\end{observation}

\begin{proof}
By observing that the role of the series (resp. parallel) nodes in the series-parallel graph is the same as the merging (resp. splitting) nodes, one can easily transform the proof of \cref{lem:s-t:concentration} into the proof of \cref{obs:s-t:concentration}.
\end{proof}

\begin{lemma}
Consider an arbitrary agent $i$, \cref{alg:rounding} returns a feasible subtree $\bT'$ such that $f_i(\bT') \leq 4H\cdot \log k \cdot \gs$ with high probability, where $H$ is the height of the decomposition tree of the series-parallel graph and $\gs$ is a guess of the optimal objective value such that the corresponding \eqref{sp:tree-LP} admits a feasible solution.
\label{lem:s-t:union}
\end{lemma}

\begin{proof}
We have the following inequalities:
\begin{align*}
&\E\left[\left(1+\frac{1}{2H}\right)^{f_i(\bT')/\gs}\right] \leq 1+\frac{1}{H} \tag*{[By \cref{obs:s-t:concentration}]} \\
\Rightarrow & \Pr\left[ \left(1+\frac{1}{2H}\right)^{f_i(\bT')/\gs} \geq k^2 \right] \leq \frac{1}{k^2}+\frac{1}{k^2 H} \tag*{[By Markov bound]} \\
\Rightarrow & \Pr\left[ f_i(\bT') \cdot \log(1+\frac{1}{2H}) \geq 2\log k \cdot \gs \right] \leq \frac{1}{k^2}+\frac{1}{k^2 H} \\
\Rightarrow & \Pr\left[ f_i(\bT') \geq 8H\log k \cdot \gs \right] \leq \frac{1}{k^2}+\frac{1}{k^2 H} \tag*{[By $\log(1+\frac{1}{2H}) \geq \frac{1}{4H}$]}
\end{align*}
\end{proof}

\begin{proofof}{\cref{thm:general:graph}}
After finishing the doubling step, we can get a guess $\gs^*$ such that $\gs^*$ can make \eqref{tree-LP} admit a feasible solution.
By \cref{lem:s-t:alg-prop}, we know that the subtree $\bT'$ returned by \cref{alg:rounding} must be a feasible subtree.
By \cref{lem:s-t:union}, we know that there is a randomized algorithm that outputs a feasible subtree such that, for each $i\in[k]$:
$$
\Pr\left[ f_i(\bT') \geq 8H\log k \cdot \gs^* \right] \leq \frac{1}{k^2}+\frac{1}{k^2 H}.
$$
By union bound, we have:
$$
\Pr\left[ \max_{i\in[k]}f_i(\bT') \leq 8H\log k \cdot \gs^* \right] \geq 1- \left(  \frac{1}{k}+\frac{1}{k H} \right).
$$
By the definition of the cost $f$ function of $\bT$ and \cref{lem:s-t:feasibility}, we know that $f_i(\bT')=c_i(\cP)$ for each $i\in[k]$.
Thus, we have:
$$
\Pr\left[ \max_{i\in[k]}c_i(\cP)=\max_{i\in[k]}f_i(\bT') \leq 8H\log k \cdot \gs^* \right] \geq 1- \left(  \frac{1}{k}+\frac{1}{k H} \right).
$$
Note that $H=2\cdot \ceil{\log n }+1 \leq 2 \log n+3$.
Plugging in, we have:
$$
\Pr\left[ \max_{i\in[k]}c_i(\cP) \leq 8 \cdot (2 \log n+3) \cdot \log k \cdot \gs^* \right] \geq 1- \left(  \frac{1}{k}+\frac{1}{k H} \right).
$$
By \cref{obs:gp:guess}, we have $\gs^* \leq 2\cdot \opt$ and thus we have:
$$
\Pr\left[ \max_{i\in[k]}c_i(\cP) \leq (32 \log n \log k+48 \log k )\cdot \opt \right] \geq 1- \left(  \frac{1}{k}+\frac{1}{k H} \right).
$$
\end{proofof}

\subsection{Hardness Result}
\label{sec:quasi:hardness}

In this section, we show the hardness result by extending the proof of $\Omega(\log^{1-\epsilon}k)$ stated in~\cite{DBLP:journals/ipl/KasperskiZ09}.
Formally, we show the following theorem (\cref{thm:quasi:hardness}).

\begin{theorem}
Any quasi-polynomial time algorithm has an approximation ratio of $\Omega(\log^{1-\epsilon}k)$ for any $\epsilon>0$ even on series-parallel graphs under the assumption of $\NP \nsubseteq \mathrm{DTIME}(n^{\poly \log n})$, where $k$ is the number of agents.
\label{thm:quasi:hardness}
\end{theorem}

\begin{proof}
    
\cref{thm:quasi:hardness} is an immediate extension of the proof of $\Omega(\log^{1-\epsilon} k)$.
In~\cite{DBLP:journals/ipl/KasperskiZ09}, they do a reduction from the 3-SAT problem.
Given a 3-SAT instance with $n$ variables, they construct an instance $\cI:=(G,\cF)$ of the robust s-t path problem, where $G$ is a series-parallel graph with $O(n^{\poly \log n})$ vertices and $\cF$ is a set of cost functions with the size of $O(\poly\log n)$. 
And they show the following key claim.

\begin{claim}[\cite{DBLP:journals/ipl/KasperskiZ09}]
Fix any $\epsilon>0$, a polynomial time algorithm approximating $\cI$ within a factor of $\log^{1-\epsilon} k$ implies a quasi-polynomial time algorithm that can decide the original 3-SAT instance.
\end{claim}

Since the size of $\cI$ is quasi-polynomial, such an algorithm shall be a quasi-polynomial time algorithm for the original 3-SAT instance.
Thus, this would imply that the 3-SAT problem admits a quasi-polynomial time algorithm which leads to a contradiction.
Because the assumption $\NP \nsubseteq \mathrm{DTIME}(n^{\poly \log n})$ assumes that NP-Complete problems do not have a quasi-polynomial time algorithm.
We show that the above claim can be stronger by the following simple observation: $n^{{\poly\log n}^{\poly \log n}}$ is still quasi-polynomial size. 
Formally, we have the following claim.

\begin{claim}
Fix any $\epsilon>0$, a quasi-polynomial time algorithm approximating $\cI$ within a factor of $\log^{1-\epsilon} k$ implies  a quasi-polynomial time algorithm that can decide the original 3-SAT instance.
\end{claim}

Suppose that there is quasi-polynomial time algorithm $\alg$ approximating $\cI$ within a factor of $\log^{1-\epsilon} k$.
By their proof, $\alg$ can be used to distinguish the original 3-SAT instance.
Since the size of $\cI$ is $O(n^{\poly \log n})$, then the running time of $\alg$ would be $O(n^{{\poly\log n}^{\poly \log n}})$ which is still in $O(n^{\poly \log n})$.
Thus, $\alg$ is a quasi-polynomial time algorithm for the 3-SAT problem that can decide any instance of 3-SAT.
This contradicts the assumption of $\NP \nsubseteq \mathrm{DTIME}(n^{\poly \log n})$.
\end{proof}

\section{Graphs with Bounded Treewidth}
\label{sec:treewidth}

In this section, we consider a special class of the directed graph which is called the graph with bounded treewidth and we describe a randomized algorithm that achieves $O(\log n \log k)$ approximation with high probability for this set of directed graphs (\cref{thm:ratio:treewidth}).
Note that this ratio leaves a logarithmic gap since there is a lower bound $\Omega(\log^{1-\epsilon} k)$ (for any $\epsilon>0$) for series-parallel graphs which have the treewidth of $2$~\cite{DBLP:journals/ipl/KasperskiZ09}.
In the following, we use $\tw(G)$ to denote the treewidth of graph $G$.

\begin{theorem}
Given any directed graph $G$ with treewidth $\tw(G)\leq \ell$, there is an algorithm that returns a $O(\log n \log k)$-approximate solution in $\poly(n) \cdot n^{O(\ell^2)}$ time with probability at least $1-(\frac{1}{k}+\frac{1}{k\log n})$ for robust $s\tto t$ path, where $n$ is the number of vertices and $k$ is the number of agents.
\label{thm:ratio:treewidth}
\end{theorem}

In the following, we first present the basic definition of the treewidth in \cref{sec:treewidth:concepts}.
And then, we discuss the intuition of our algorithm and give the complete description of our algorithm in \cref{sec:treewidth:framework}.
We show the detailed reduction in \cref{sec:reduction}.
Finally, we prove \cref{thm:ratio:treewidth} in \cref{sec:treewidth-proof}.

\subsection{Basic Concepts}
\label{sec:treewidth:concepts}

The treewidth of a directed graph is defined as the treewidth of its underlying undirected graph.
The underlying undirected graph of a directed graph is one with the same set of vertices, but a different set of edges. 
In particular, between any pair of vertices $v$ and $v'$, if the directed graph has an edge $v\to v'$ or an edge $v'\to v$, the underlying undirected graph includes an edge $(v,v')$.
Treewidth has several different equivalent definitions. 
In this paper, we use its definition based on tree decomposition.

\begin{definition}[Tree Decomposition]
Let $G:=(V,E)$ be an undirected graph. 
Let $\bT:=(\bV,\bE)$
be a tree whose nodes are subsets of vertices of $E$, i.e., $\bu \subseteq V$ for all $\bu \in \bV$.
Then, $\bT$ is a tree decomposition of $G$ if $\bT$ has the following properties:
\begin{enumerate}[label=(P\arabic*),leftmargin=*,align=left]
    \item {\bf (Completeness)} Each edge $(p,q)\in E$ is contained in a node, i.e., some nodes of $\bT$ contain both $p$ and $q$.
    \label{prop:completeness}
    \item {\bf (Connectivity)} For an arbitrary vertex $p\in V$, all nodes of $\bT$ that contain $p$ form a connected component.
    \label{prop:connectivity}
\end{enumerate}
\label{def:tree-decomposition}
\end{definition}

The {\em width} of a tree decomposition is the size of its largest node minus one\footnote{The size of the largest set is diminished by one in order to make the treewidth of a tree equal to one.}.
The treewidth of an undirected graph $G$ is the minimum width among all its possible tree decompositions of $G$. 
In~\cite{DBLP:conf/wg/Bodlaender88}, they proved that a tree decomposition can be adjusted to have a height of logarithmic level while sacrificing a constant factor on its width.
Formally, this can be captured by \cref{lem:tree_decomp_height}.
This property is crucial for our algorithm as the running time and approximation ratio of our algorithm depend on the height of the tree decomposition.

\begin{lemma}[\cite{DBLP:conf/wg/Bodlaender88}]
Given an arbitrary undirected graph $G(V,E)$ with $\abs{V}=n$ and $\tw(G)\leq \ell$,
we can compute a tree decomposition $\bT$ of $G$ in linear time such that (\rom{1}) $\bT$ is a binary tree with height $\leq 2\ceil{\log_{\frac{5}{4}} 2n}$; (\rom{2}) the width of $\bT$ is at most $3\ell+2$.
\label{lem:tree_decomp_height}
\end{lemma}

\subsection{Algorithmic Framework}
\label{sec:treewidth:framework}

In this section, we give our algorithmic framework.
Generally, we aim to reduce our problem to {\em the tree labeling problem} which was proposed by Dinitz et al.~\cite{DBLP:journals/corr/abs-2302-11475} very recently.
We restate the definition and the result of the tree labeling problem in \cref{sec:tree-labeling}.
At first glance, the tree labeling problem seems quite dissimilar to our problem. 
Therefore, in \cref{sec:reduction-intuition}, we provide some intuition for the reduction. 
In order to ensure the correctness of the reduction, we additionally need to preprocess the graph and utilize some other tools. 
Thus, we give a complete description of our algorithm in \cref{sec:complete-algorithm}.

\subsubsection{The Tree-labeling Problem}
\label{sec:tree-labeling}


We are given a binary tree $\bT:=(\bV,\bE)$ rooted at $\br\in \bV$.
For each node $\bv\in\bV$, there is a finite set $L_{\bv}$ of labels for $\bv$.
Let $L:=\bigcup_{\bv\in\bV}L_{\bv}$ be the set of all possible labels.
The output is a label assignment $\cL:=(l_{\bv}\in L_{\bv})_{\bv\in\bV}$ of the node set $\bV$, that satisfies the following consistency and cost constraints.
\begin{itemize}
    \item {\bf (Consistency Constraints)}
    For every internal node $\bv$ of $\bT$ with two children $\bu$ and $\bw$ ($\bu$ or $\bw$ is possibly empty), we are given a set $\Gamma(\bv)\subseteq L_{\bu} \times L_{\bv} \times L_{\bw}$.
    A valid labeling $\cL=(l_{\bv}\in L_{\bv})_{\bv\in\bV}$ must satisfy $(l_{\bu},l_{\bv},l_{\bw})\in\Gamma(\bv)$ for every internal node $\bv$.
    \item {\bf (Cost Constraints)} 
    There are $k$ additive cost functions $f_1,\ldots,f_k$ defined over the labels, i.e., for each $i\in[k]$, $f_i:L\to \R_{\geq 0}$.
    For each $i\in[k]$, a valid labeling $\cL$ needs to satisfy $f_i(\cL):=\sum_{\bv\in\bV}f_i(l_{\bv})\leq 1$.
\end{itemize}

A label assignment $\cL:=(l_{\bv}\in L_{\bv})_{\bv\in\bV}$ is called {\em consistent} if it satisfies the consistency constraints; it is {\em valid} if it satisfies both the consistency and cost constraints.
Let $H$ be the height of $\bT$ and let $\Delta:=\max_{\bv\in\bV}\abs{L_{\bv}}$ be the maximum size of any label set.
Let $n$ be the number of nodes in $\bT$.
In~\cite{DBLP:journals/corr/abs-2302-11475}, they showed the following result.

\begin{lemma}[\cite{DBLP:journals/corr/abs-2302-11475}]
Given a tree labeling instance such that the instance admits a valid label assignment.
There is a randomized algorithm that in time $\poly(n) \cdot \Delta^{O(H)}$ outputs a consistent labeling $\cL$ such that for every $i\in[k]$, we have
$$
\E\left[\exp\left( \ln(1+\frac{1}{2H})\cdot f_i(\cL) \right)\right] \leq 1+\frac{1}{H}.
$$
\label{lem:tree-labeling-ratio}
\end{lemma}

\subsubsection{Construction Intuition}
\label{sec:reduction-intuition}

In this section, we give some intuition of our reduction.
We focus on the high-level idea of the instance construction and thus some parts may have inaccurate descriptions.
The formal description can be found in \cref{sec:reduction}.

Generally, given any directed graph with bounded treewidth, we aim to construct a tree-labeling instance such that the solution to the constructed tree-labeling instance can be converted into an $s \tto t$ path of the original graph with some cost-preserved property.
The first step of the reduction is to build a binary tree for the tree-labeling instance. 
By \cref{lem:tree_decomp_height}, we know that there exists a tree decomposition $\bT:=(\bV,\bE)$ with logarithmic depth for any directed graph with bounded treewidth.
We directly use $\bT$ as the binary tree in the tree-labeling instance.
To build the reduction, we need to address two issues. 
The first issue is the meaning of labels on nodes, and the second issue is how to impose constraints (consistency constraints) on the labels such that a feasible label assignment can be successfully transformed into an $s \tto t$ path. 
In this section, we can temporarily ignore the cost constraints of the tree labeling problem, which we will address in \cref{sec:complete-algorithm}.
Thus, we shall focus on how to design the consistency constraints so that a label assignment is able to be converted into an $s \tto t$ path if the assignment satisfies the consistency constraints.
In the following, we briefly discuss our ideas for resolving these two issues.

\paragraph{Label Construction.}
Note that an assignment of labels needs to be translated into an $s \tto t$ path, which means that the assignment needs to distinguish which edges are on this $s \tto t$ path. 
To achieve this, we assign a ``choosing label'' (which is either $\n$ or $\p$) to the edges contained in node $\bu$ to indicate whether the edge is selected or not. 
The completeness (\ref{prop:completeness}) of a tree decomposition ensures that we can do this successfully, as \ref{prop:completeness} guarantees that every edge is included in some node. 
Thus, we can clearly determine whether an edge is on the final $s \tto t$ path.
However, having only one choosing label is not enough because without the other labels, we cannot guarantee that an assignment of labels can be transformed into an $s \tto t$ path, that is, we cannot ensure the connectivity between $s$ and $t$. 
To do so, we introduce a ``connectivity label'' (which is either $\n$ or $\p$) for each vertex pair $(a,b)$ in node $\bu$ to indicate whether $a$ and $b$ are connected in the subtree rooted at node $\bu$, i.e., whether the selected edges in some nodes in the subtree rooted at node $\bu$ can connect $a$ and $b$. 
Note that some nodes may not contain source $s$ and sink $t$, so we add $s$ and $t$ to each node in $\bT$ additionally. 
After adding $s,t$ to all nodes, the resulting tree is still a tree decomposition, and the width of $\bT$ increases by at most $2$. 
Now, fix any node $\bv$ in $\bT$, edges contained in $\bv$ have a choosing label and each vertex pair in $\bv$ has a connectivity label. 
Their combination constitutes a possible label for this node.

As one may observe, the above reduction might have some potential issues, because the tree-labeling problem involves only a ``local'' constraint (consistency constraint). 
This means that the connectivity between a vertex pair $(a,b)$ in node $\bu$ is only influenced by the connectivity between $a$ and $b$ in its child nodes.
This might prevent the connectivity between $a$ and $b$ in node $\bu$ from reflecting the connectivity between $a$ and $b$ in the subtree rooted at node $\bu$.
We show that this issue is fixable by setting up some appropriate local constraints.
Namely, for an arbitrary vertex pair $(a,b)$ in $\bu$, we can obtain the true connectivity of $a,b$ in the subtree rooted at $\bu$ by only using the connectivity information from $\bu$ and $\bu$'s child nodes.

\paragraph{Constraint Construction.}
Now we fix any node $\bu$ in $\bT$ and focus on an arbitrary vertex pair $(a,b)$ in $\bu$.
We aim to seek some conditions under which the connectivity label of $(a,b)$ can be set to $\p$ in $\bu$ such that the true connectivity of $a,b$ in the subtree rooted at $\bu$ is stored by $(a,b)$'s connectivity label.
Intuitively, if vertex $c$ and $d$ have a connectivity label of $\p$ in some node $\bv$, then there exists a ``connected segment'' from $c$ to $d$ in the subtree rooted at $\bv$. 
Thus, the most basic condition is that if the connected segment in $\bu$ and its children node are able to form a connected segment from $a$ to $b$, then the connectivity label of $(a,b)$ is $\p$.
For example, if there is a connected segment $c \to \cdots \to d$ in $\bu$, and connected segments $d\to \cdots \to b$, $a\to \cdots \to c$ in $\bu$'s child nodes, then these three connected segments are able to be merged into a path from $a$ to $b$ and thus the connectivity label is $\p$.
Surprisingly, we show that such a simple and basic condition is sufficient.
Namely, the connectivity information contained in node $\bu$ and its child nodes can make $a$ and $b$ connected if and only if they are connected in the subtree rooted at $\bu$.
We formally show this claim in \cref{clm:reduction} which is included in the proof of \cref{lem:reduction:key}.
The correctness of the above simple condition heavily relies on the connectivity of a tree decomposition (\ref{prop:connectivity}).
The connectivity of a tree decomposition states that all nodes containing the same vertex form a connected subtree, which intuitively ensures that the connectivity between vertices can be continuously propagated between nodes.
For example, consider the node $\bu$ and a vertex pair $(a,b)$ in $\bu$.
Let $\bv_1$ be the child node of $\bu$ and let $\bv_2$ and $\bv_3$ be the children node of $\bv_1$.
If $a,c$ are connected in $\bv_2$ and $c,d$ are connected in $\bv_3$, then the connectivity label of $a,b$ must be $\p$.
This is due to \ref{prop:connectivity} of a tree decomposition, i.e., if $\bu$ includes $a,b$, $\bv_2$ includes $a$, and $\bv_3$ includes $b$, then node $\bv_1$ must also include $a,b$.
Thus, the connectivity information of $a,b$ will not be interrupted by $\bv_1$.

\subsubsection{The Complete Algorithm}
\label{sec:complete-algorithm}

In this section, we present the complete algorithm (\cref{alg:treewidth}) for graphs with bounded treewidth.
Recall that we aim to use the algorithm of the tree labeling problem.
The complete algorithm mainly consists of the following five steps.

\begin{enumerate}[label=(S\arabic*),leftmargin=*,align=left]
  \item Doubling Step (lines \ref{line:binary-search-1}-\ref{line:binary-search-2} of \cref{alg:treewidth}): \\
  To show \cref{thm:ratio:treewidth} by \cref{lem:tree-labeling-ratio}, we need to find an ``appropriate'' guess $\gs$ of the optimal objective value which is addressed in this step.
  Namely, in this step, we aim to find a $\gs$ such that $\gs \leq c\cdot \opt$, where $c$ is some small constant and $\opt$ is the optimal objective value of our problem.
  To this end, we employ the standard doubling technique and start with an upper bound of the optimal solution, e.g., $\max_{i\in[k]}\sum_{i\in [k]}c_i(e)$ which we use in \cref{alg:treewidth}.
  After having a $\gs$, we shall cut the large edge (see the second step below) to check whether the constructed tree-labeling instance has a valid label assignment.
  If so, we continue to decrease the value of $\gs$ by dividing $2$; otherwise, we just stop.
  Finally, \cref{alg:treewidth} ensures that $\gs \leq 2\cdot \opt$ (see \cref{obs:guess}).
  \item Cut Large Edges (lines \ref{line:cut-edge-1}-\ref{line:cut-edge-2} of \cref{alg:treewidth}):\\
  In this step, we address the cost constraints of the tree-labeling problem.
  Now we assume that we obtain a guess $\gs$ such that $\gs \leq 2\cdot \opt$.
  Then, we preprocess the input graph by removing those edges with a large cost.
  Specifically, an edge $e$ should be discarded if there exists an agent $i\in[k]$ such that $c_i(e) > \gs$.
  Because no solution with the objective value of $\gs$ can use these large edges.
  \item Tree Decomposition Computing (lines \ref{line:tree-decomposition-1}-\ref{line:tree-decomposition-2} of \cref{alg:treewidth}):\\
  In this step, we first construct an underlying undirected graph of the given directed graph.
  And then, we employ \cref{lem:tree_decomp_height} to compute a desired tree decomposition.
  Finally, we add the source $s$ and $t$ to all nodes in the computed tree decomposition for the purpose of the reduction.
  \item Tree-labeling Solving (lines \ref{line:tree-labeling-1}-\ref{line:tree-labeling-2} of \cref{alg:treewidth}):\\
  In this step, we construct a tree-labeling instance using the tree decomposition from the second step by the methods stated in \cref{sec:reduction}.
  After that, we employ the algorithm in~\cite{DBLP:journals/corr/abs-2302-11475} to obtain a feasible label assignment.
  \item $s \tto t$ Path Construction (lines \ref{line:stpath-1}-\ref{line:stpath-2} of \cref{alg:treewidth}): \\
  In this step, we convert the obtained feasible label assignment to an $s \tto t$ path by \cref{lem:reduction:map}.
  Note that the obtained feasible label assignment may be mapped into a subgraph instead of a single $s \tto t$ path.
  But by our construction method, we can ensure that such a subgraph must contain at least one $s \tto t$ path (\cref{lem:reduction:map}).
  In this case, we just arbitrarily pick an $s \tto t$ path in the obtained subgraph.
  This can only decrease the cost of the solution.
\end{enumerate}

\begin{algorithm}[htb] 
\caption{The Complete Algorithm for Graphs with Bounded Treewidth}
\label{alg:treewidth}
\begin{algorithmic}[1]
\Require A directed graph $G:=(V,E)$ with $k$ cost functions $c_i:2^{E}\to \R_{\geq 0},i\in[k]$.
\Ensure An $s \tto t$ path $\cP\subseteq E$.
\State $\flag \leftarrow \true$; $\gs \leftarrow \max_{i\in[k]}\sum_{e\in E}c_i(e)$.
\While{$\flag = \true$}
\label{line:binary-search-1}
\State $E'\leftarrow\set{e\in E \mid \exists i\in[k]\text{ s.t. }c_i(e)>\gs}$.
\label{line:cut-edge-1}
\State $E\leftarrow E\setminus E'$; $G\leftarrow (V,E')$.
\State Normalize all cost functions, i.e., $c_i(e) \leftarrow \frac{c_i(e)}{\gs}$ for all $i\in[k]$ and $e\in E$.
\label{line:cut-edge-2}
\State Let $G'$ be the underlying undirected graph of $G$.
\label{line:tree-decomposition-1}
\State Compute a tree decomposition $\bT:=(\bV,\bE)$ of $G'$ by \cref{lem:tree_decomp_height}.
\State Add $s$ and $t$ to all nodes in $\bV$.
\label{line:tree-decomposition-2}
\State Construct a tree-labeling instance by the method in \cref{sec:reduction}.
\label{line:tree-labeling-1}
\State Solve the tree-labeling instance using \cite{DBLP:journals/corr/abs-2302-11475}.
\If{the instance has a valid label assignment}
\State Let $\cL$ be returned label assignment.
\State $\flag \leftarrow \true$; $\gs \leftarrow \frac{\gs}{2}$.
\Else
\State $\flag \leftarrow \false$.
\EndIf
\label{line:tree-labeling-2}
\EndWhile
\label{line:binary-search-2}
\State Convert the $\cL$ to a subgraph $G''$ of $G$ by \cref{lem:reduction:map}.
\label{line:stpath-1}
\State Pick an arbitrary $s \tto t$ path $\cP$ in $G''$.
\label{line:stpath-2}
\State \Return $\cP$.
\end{algorithmic}
\end{algorithm}

\subsection{Tree-labeling Instance Construction}
\label{sec:reduction}

Given an arbitrary node $\bv\in\bV$, we also use $\bv$ to denote the vertices included in node $\bv$, i.e., $\bv\subseteq V$.
Let $E_{\bv}\subseteq E$ be the set of edges such that, for any edge $(a,b)$ in $E_{\bv}$, node $\bv$ is the highest node that contains $(a,b)$. 
Note that an edge $(a,b)$ may be included in more than one node in $\bT$ but the highest node that includes $(a,b)$ is unique.
For any node $\bv\in\bV$, we have two types of labels: choosing label and connectivity label.
For each edge $(a,b)$ (or $e$) in $E_{\bv}$, the choosing label $\chng(a,b)=\p$ or (or $\chng(e)=\p$) indicates that the edge is chosen in current label assignment; otherwise, the edge is not chosen.
For each pair of vertices $(p,q)$ in $\bv$, the connectivity label $\conn(p,q)=\p$ indicates that there is a $\cP\subseteq E$ path from $p$ to $q$ such that every edge in $\cP$ is chosen in some nodes of the subtree rooted at node $\bv$; otherwise, $p$ and $q$ are not connected.
Note that $\conn(p,q)$ and $\conn(q,p)$ are two different labels since $G$ is a directed graph.
Let $L_{\bv}$ be the set of all possible labels and $l_{\bv}\in L_{\bv}$ be a specific label of node $\bv$.
We remark that the size of $L_{\bv}$ is related to the treewidth of $G$.
Since the treewidth of $G$ is a constant, $\abs{L_{\bv}}$ is also constant.
See the proof of \cref{thm:ratio:treewidth} for details.

To ensure the feasibility of label assignments for obtaining an $s \tto t$ path in $\bT$, arbitrarily picking labels for each node is not a viable solution. 
Instead, we define a local constraint that applies to every adjacent set of three nodes $\bu,\bv,\bw$ in $\bT$, where $\bu$ and $\bw$ are the child nodes of $\bv$. 
The purpose of this local constraint is to guarantee that all label assignments are capable of producing an $s \tto t$ path.
For every node $\bv$ and its two children $\bu$ and $\bw$, let $\texttt{CP}(\bv):=L_{\bu} \times L_{\bv} \times L_{\bw}$ be the set of all possible label combinations of these three nodes, i.e., $\texttt{CP}(\bv)$ is the Cartesian product of $L_{\bu},L_{\bv},L_{\bw}$.
Note that $\bu$ or $\bw$ may not exist.
In this case, we refer to $\bu$ or $\bw$ as empty nodes and $\texttt{CP}(\bv)$ is defined as the $L_{\bv}\times L_{\bw}$ (or $L_{\bu} \times L_{\bv}$ or $L_{\bv}$).

\begin{definition}[Feasible Label Assignment]
A label assignment $\cL:=(l_{\bv})_{\bv\in\bV}$ is a feasible label assignment if, for each $\bv\in\bV$ and its two children $\bu$ and $\bw$ ($\bu$ and $\bw$ maybe empty nodes), $l_{\bv}\in \textup{\texttt{CP}}(\bv)$ satisfies the following three constraints:
\begin{enumerate}[label=(C\arabic*),leftmargin=*,align=left]
    \item {\bf (Choosing Constraints)}
    For each edge $(a,b)\in E_{\bv}$, if $(a,b)$ is chosen, then $a$ and $b$ are connected and vice versa. 
    Namely, $\chng(a,b)=\p$ if and only if $\conn(a,b)=\p$.
    \label{cons:choosing}
    \item {\bf (Connectivity Constraints)}
    For each vertex pair $(p,q)$ in $\bv$, vertex $p$ and vertex $q$ are connected (i.e. $\conn(p,q)=\p$) if and only if the following statement is true:
    there is a vertex sequence $(p, v_1,\ldots,v_d,q)$ such that every two adjacent vertices $(a,b)$ in the sequence are connected in some nodes in $\bu,\bv,\bw$, i.e., $\conn(a,b)=\p$ in some nodes in $\bu,\bv,\bw$ for each pair of adjacent vertices.
    \label{cons:connectivity}
    \item {\bf (Feasibility Constraints)}
    If $\bv$ is the root of $\bT$, then the source $s$ and sink $t$ are connected, i.e., $\conn(s,t)=\p$ must be true in the root.
    \label{cons:feasibility}
\end{enumerate}
\label{def:feasible-label}
\end{definition}

Given a feasible label assignment $\cL$, an edge $(p,q)\in E$ is {\em chosen} by $\cL$ if $(p,q)$'s choosing label is $\p$ in $\cL$.
\ref{cons:choosing} defines the connectivity of each edge in $E$.
If an edge $(a,b)$ is chosen by a label assignment, then vertex $a$ and $b$ are connected.
\ref{cons:connectivity} is the most important constraint which defines the connectivity of each pair of vertices in $\bv$.
In the case where $\bv$ is a leaf node, $\bu$ and $\bw$ would be empty nodes and thus this constraint is equivalent to \ref{cons:choosing}.
In the case where $\bv$ is not a leaf node, an arbitrary vertices pair $(a,b)$ in $\bv$ are connected if the connected segments in $\bu,\bv,\bw$ can be merged into a path from $a$ to $b$.
In \cref{lem:reduction:key}, we show that such a local constraint is sufficient to describe the connectivity of vertex $a$ and $b$ in the subtree rooted at $\bv$.
\ref{cons:feasibility} ensures that a feasible label assignment must contain an $s \tto t$ path, i.e., source $s$ and sink $t$ are connected.

\begin{figure}[htb]
    \centering
    \includegraphics[width=15cm]{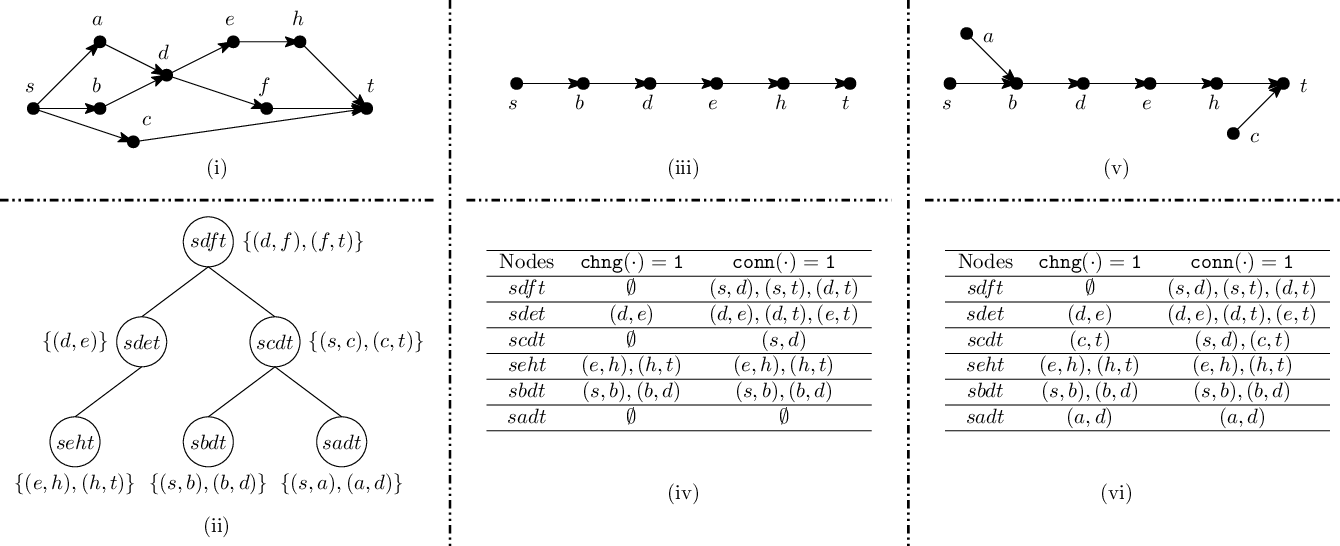}
    \caption{An example of the reduction. The subfigure (\rom{1}) is the given directed graph. Then, we compute a tree decomposition with the logarithmic depth and add $s$ and $t$ to all nodes, which is shown in subfigure (\rom{2}). The edge set next to each node in subfigure (\rom{2}) is its corresponding $E_{\bv}$. For example, the edge set $E_{\br}$ for root node $\br$ is $\set{(d,f),(f,t)}$ since $\br$ is the highest node that contains edge $(d,f)$ and $(f,t)$. Subfigure (\rom{3}) is an $s \tto t$ path of the given directed graph and subfigure (\rom{4}) is the corresponding label assignment of the $s \tto t$ path in (\rom{3}) in which we only list these labels with the value of $\p$. The complete label of each node is obtained by merging these single labels, e.g., for the root $\br$, $l_{\br}$ consists of $14$ bits ($2$ choosing labels and $12$ connectivity labels). In these $14$ bits, only $\conn(s,d),\conn(s,t),\conn(d,t)$ has a value of $\p$ and all the remaining $11$ labels have a value of $\n$. Subfigure (\rom{5}) shows another example and subfigure (\rom{6}) is its corresponding label assignment. By our construction, it is possible that a feasible label assignment corresponds to a subgraph instead of a single $s \tto t$ path.}
    \label{fig:treewidth_reduction_new}
\end{figure}

Consider an arbitrary feasible label assignment $\cL$, then $\cL$ has the following crucial property (\cref{lem:reduction:key}) by our definition.
We shall use this property later to show that any feasible label assignment can be converted into an $s \tto t$ path of the original graph (\cref{lem:reduction:map}).
It is worth noting that an $s \tto t$ path corresponds to a unique feasible label assignment, but a feasible label assignment may contain multiple $s \tto t$ paths.
Before presenting the formal proof, we give an example in \cref{fig:treewidth_reduction_new}.

\begin{lemma}
Given an arbitrary feasible label assignment $\cL:=(l_{\bv})_{\bv\in\bV}$, consider an arbitrary node $\bv\in\bV$.
For any vertices pair $(p,q)$ in $\bv$, there is a path $\cP\subseteq E$ from $p$ to $q$ such that every edge in $\cP$ is chosen in some nodes in the subtree rooted at $\bv$ if and only if $(a,b)$ has a connectivity label of $\p$ in $\bv$.
\label{lem:reduction:key}
\end{lemma}

\begin{proof}
If $\bv$ is a leaf node, the lemma trivially holds since the subtree rooted at $\bv$ is just $\bv$.
In this case, $(p,q)$ has the connectivity label of $\p$ if and only if edge $(p,q)$ is chosen by \ref{cons:choosing}.
Now we assume that $\bv$ is not a leaf node.
In the following, we show the lemma from both directions.

\paragraph{($\Rightarrow$)}
In this direction, we show that if there is a path $\cP\subseteq E$ from $p$ to $q$ such that every edge in $\cP$ is chosen in some nodes in the subtree rooted at $\bv$, then $\conn(a,b)=\p$ in $\bv$.
Before presenting the proof, we first define some notations.
We use the notation $a \rightsquigarrow b$ to represent that there exists a path from vertex $a$ to vertex $b$.
Let $\bT':=(\bV',\bE')$ be the subtree rooted at $\bv$.
Given any node $\bu$ in $\bT'$, let $\edge(\bu)\subseteq \cP$ be the set of edges such that (\rom{1}) each edge in $\edge(\bu)$ is included in some nodes in the subtree rooted at $\bu$; (\rom{2}) each edge in $\edge(\bu)$ is included in the path $\cP$.
Note that $\edge(\bu)$ can be considered as a set of subpaths of $\cP$ for all $\bu\in\bV'$.
We use $\edge(\bu) \cup \edge(\bw)$ to denote the set of subpaths after merging, e.g., if there is a subpath $v\rightsquigarrow v'$ in $\edge(\bu)$ and there is a subpath $v'\rightsquigarrow v''$ in $\edge(\bw)$, then these two subpaths are merged into a larger one $v\rightsquigarrow v''$ in $\edge(\bu) \cup \edge(\bw)$.
To show the correctness of this direction, we prove the following stronger claim.
Suppose \cref{clm:reduction} is correct, then the correctness of the lemma directly follows by applying \cref{clm:reduction} to node $\bv$.

\begin{claim}
Consider an arbitrary node in $\ba\in\bV'$.
Then, node $\ba$ stores the connectivity information of all subpaths in $\edge(\ba)$, i.e., for each subpath $v\rightsquigarrow v'$ in $\edge(\ba)$, $\conn(v,v')=\p$ in node $\ba$.
\label{clm:reduction}
\end{claim}

\begin{proof}
We prove the claim by induction in which we prove from the bottom to the top of tree $\bT'$.
When $\ba$ is a leaf node, the claim trivially holds by \ref{cons:choosing}.
Let $\bb$ and $\bc$ be $\ba$'s two child nodes.
Now we suppose that $\bb$ and $\bc$ stores the connectivity information of all subpaths in $\edge(\bb)$ and $\edge(\bc)$, respectively.
In the following, we prove the claim by applying \ref{cons:connectivity} to $\ba,\bb,\bc$.
Let $f\rightsquigarrow h$ be an arbitrary subpath in $\edge(\ba)\cup\edge(\bb)\cup\edge(\bc)$.
We claim that $\conn(f,h)=\p$ must be true in $\ba$.
The subpath $f\rightsquigarrow h$ may come from one of the following three cases:
(\rom{1}) $f\rightsquigarrow h$ is in $\edge(\ba)$;
(\rom{2}) $f\rightsquigarrow h$ is in $\edge(\bb)$ or $\edge(\bc)$;
(\rom{3}) $f\rightsquigarrow h$ is a new subpath after merging some subpaths in $\edge(\ba),\edge(\bb),\edge(\bc)$.
If case (\rom{1}) is true, then $f$ and $h$ have a connectivity label of $\p$ in $\ba$.
The correctness of case (\rom{2}) and (\rom{3}) stems from the connectivity (\ref{prop:connectivity}) of a tree decomposition and our local constraints defined in \cref{def:feasible-label}.

\paragraph{Case (\rom{2}).}
Note that if node $\ba$ contains $f$ and $h$, then $\conn(f,h)=\p$ must be true by \ref{cons:connectivity}.
Thus, to show the claim, it is sufficient to show that $f,h\in\ba$.
To this end, we only need to show that there exists a node $\bx$ (resp. $\by$) such that (\rom{1}) $\bx$ (resp. $\by$) contains $f$ (resp. $h$); (\rom{2}) $\bx$ (resp. $\by$) is in the subtree rooted at $\bv$; (\rom{3}) $\bx$ (resp. $\by$) is not in the subtree rooted at $\ba$.
If we find such a $\bx$ and $\by$, $\ba$ must contain $f$ and $h$ by the connectivity (\ref{prop:connectivity}) of a tree decomposition.
If $f$ (resp. $h$) is the source $s$ or $p$ (resp. the sink $t$ or $q$), then the root $\bv$ of subtree $\bT'$ is such a node.
Now, we suppose that $f$ is not the $s,p$ and $h$ is not the $t,q$.
This implies that there must exist an edge $(v, f)$ and an edge $(h,v')$ since both $f,h$ are in the path from $p$ to $q$.
Let $\bs$ (resp. $\bs'$) be the node that contains $(v, f)$ (resp. $(h,v')$).
Note that $\bs$ and $\bs'$ must not be in the subtree rooted at $\ba$; otherwise, the subpaths in $\edge(\bb)$ or $\edge(\bc)$ would be $v\rightsquigarrow v'$ instead of $f\rightsquigarrow h$.
Thus, $\bs$ and $\bs'$ are the nodes that we need.

\paragraph{Case (\rom{3}).}
To simplify the analysis, we assume that $f\rightsquigarrow h$ comes from the following merging: subpath $f\rightsquigarrow a$ and subpath $b\rightsquigarrow h$ are merged via subpath $a\rightsquigarrow b$.
For other types of merging, the proof is the same.
Note that $\conn(f,a)=\p$, $\conn(b,h)=\p$ and $\conn(a,b)=\p$ must be true in some nodes in $\ba,\bb,\bc$.
In the case where $f,h\in\ba$, then we are done.
To show that $f,h\notin\ba$ is impossible, we use the same proof stated in Case (\rom{2}).
Namely, it is sufficient to show that there exists a node $\bx$ (resp. $\by$) such that (\rom{1}) $\bx$ (resp. $\by$) contains $f$ (resp. $h$); (\rom{2}) $\bx$ (resp. $\by$) is in the subtree rooted at $\bv$; (\rom{3}) $\bx$ (resp. $\by$) is not in the subtree rooted at $\ba$.
If we find such nodes, this would imply that $f,h\in\ba$ must be true.
The remaining proof is the same as Case (\rom{2})'s proof.
\end{proof}

\paragraph{($\Leftarrow$)}
To show this direction, we only need to find a path $\cP\subseteq E$ from $p$ to $q$ such that every edge in $\cP$ is chosen in some nodes in the subtree rooted at $\bv$.
To this end, we recursively apply \cref{def:feasible-label} from the top to the bottom of the subtree rooted at $\bv$.
We begin from the node $\bv$.
Since $(p,q)$ are connected in $\bv$, there must exist a vertex sequence $\sigma:=(p,v_1,\ldots,v_d,q)$ such that every two adjacent vertices $(a,b)$ in the sequence are connected in some nodes in $\bu,\bv,\bw$ by \cref{def:feasible-label}.
Let $\X$ be all adjacent vertex pairs in this vertex sequence.
Let $\X_{\bu},\X_{\bv},\X_{\bw}$ be the adjacent vertex pairs that appear in $\bu,\bv,\bw$, respectively.
If a vertex pair is included in both $\X_{\bv}$ and $\X_{\bu}$ (or both $\X_{\bv}$ and $\X_{\bw}$), then remove such a vertex pair from $\X_{\bv}$.
If a vertex pair is included in both $\X_{\bu}$ and $\X_{\bw}$, then remove such a vertex pair arbitrarily from $\X_{\bu}$ or $\X_{\bw}$.
Now $\X_{\bu},\X_{\bv},\X_{\bw}$ is a partition of the vertex pair $\X$.
We claim that each vertex pair in $\X_{\bv}$ must be an edge, i.e., for each $(a,b)$ in $\X_{\bv}$, there is an edge from $a$ to $b$.
Note that both $a$ and $b$ are not included in $\bu$ and $\bw$.
By the connectivity (\ref{prop:connectivity}) of the tree decomposition, we know that only $\bv$ includes $a$ and $b$ in the subtree rooted at $\bv$.
Thus, the connectivity of $a$ and $b$ can only come from the case where $(a,b)\in E_{\bv}$ and edge $(a,b)$ is chosen.
We found some chosen edges in $\cP$ which is $\X_{\bv}$.
Note that the vertex sequence $(p,v_1,\ldots,v_d,q)$ is decomposed into several subsequences by removing all pairs in $\X_{\bv}$.
Let $\sigma_1,\ldots,\sigma_{\kappa}$ be the resulted subsequences.
Each $\sigma_i\in\set{\sigma_1,\ldots,\sigma_{\kappa}}$ can be regarded as a connected subpath from the first to the last vertex of $\sigma_i$.
We redefine $\X_{\bu}$ and $\X_{\bv}$ as a partition of $\set{\sigma_1,\ldots,\sigma_{\kappa}}$.

Now we apply \cref{def:feasible-label} to each subsequence included in the node $\bu$ and $\bw$, respectively.
For example, let $\sigma_i:=c\rightsquigarrow d$ be an arbitrary subsequence in $\X_{\bu}$.
By \cref{def:feasible-label}, there must exist a vertex sequence $(c,\ldots,d)$ such that every two adjacent vertex pairs in the sequence are connected in some nodes in $\bu$ and its two child nodes.
Then, we apply the same proof to ensure that all vertex pairs in $\X_{\bu}$ are edges.
In this way, we get another set of edges in $\cP$.
After we recursively apply the above proof to all nodes in the subtree rooted at $\bv$, it is easy to see that $\bigcup_{\bs}X_{\bs}$ would contain all edges in $\cP$.
\end{proof}

\begin{lemma}
A feasible label assignment $\cL$ corresponds to a unique subgraph $G'$ of $G$ that contains at least one $s \tto t$ path. 
Moreover, an $s \tto t$ path $\cP$ corresponds to a unique feasible label assignment.
\label{lem:reduction:map}
\end{lemma}

\begin{proof}
We show the lemma from both directions.

\paragraph{From $\cL$ to $G'$.}
Given a feasible label assignment $\cL$, let $E'\subseteq E$ be the set of edges that have the choosing label of $\p$.
Since each edge appears exactly in one node, it is impossible that, for any $e\in E$, $e$ has the choosing label of $\p$ in some nodes of $\bT$ but it has the choosing label of $\n$ in some other nodes.
Let $G[E']$ be the induced subgraph of the edge set $E'$.
$G[E']$ is defined as the corresponding subgraph of the feasible label assignment $\cL$.
Note that given a $\cL$, $G[E']$ is unique since $\cL$ would fix the choosing label of all edges in $E$.
Now, we claim that $G[E']$ must contain at least one $s \tto t$ path. 
By \ref{cons:feasibility}, we know that the connectivity of $s$ and $t$ is $\p$ in the root $\br$ of tree $\bT$.
Then, we apply \cref{lem:reduction:key} to the root $\br$.
We conclude that there is a path $\cP\subseteq E$ from $s$ to $t$ such that every edge in $\cP$ is chosen in some nodes.

\paragraph{From $\cP$ to $\cL$.}
Consider an arbitrary $s \tto t$ path $\cP=(s,v_1,\ldots,v_d,t)$ of $G$.
To get a feasible label assignment based on $\cP$, we first set the choosing label of each edge in $\cP$ as $\p$ and all other edges have the choosing label $\n$.
And then, we start to assign the connectivity label of each vertex pair in each node.
To this end, we start from the leaf nodes of $\bT$ and assign labels level by level according to \cref{def:feasible-label}.
In each node, we first check all edges with the choosing label of $\p$, and set the connectivity label as $\p$ for those edges.
This makes sure that the assignment satisfies \ref{cons:choosing}.
If the current node is not a leaf node, we set the connectivity label of a vertex pair $(p,q)$ as $\p$ if there exists a vertex sequence $(p,v_1,\ldots,v_d,q)$ such that every two adjacent vertices in the sequence are connected in some nodes in the current node and its child nodes.
This makes sure that the assignment satisfies \ref{cons:connectivity}.
Finally, by running the above label assignment procedure, the source $s$ and sink $t$ must have a connectivity label of $\p$ by \cref{lem:reduction:key}.
This makes sure that the assignment satisfies \ref{cons:feasibility}.
Thus, we obtain a feasible label assignment.
Such a feasible label assignment is unique because the given $s \tto t$ path determines the choosing label of all edges and the connectivity labels are decided by both choosing labels and the structure of the tree decomposition.
\end{proof}

Given an arbitrary node $\bu\in\bV$, it now remains to assign a cost to each label in $L_{\bu}$.
Let $\mathsf{L}$ be the set of all feasible label assignments, i.e., each $\cL:=(l_{\bv})_{\bv\in\bV}\in \mathsf{L}$ is a feasible label assignment.
Recall that we have $k$ cost functions $c_i:2^E \to \R_{\geq 0}, i\in[k]$ defined over the edges of the original directed graph.
For each cost function $c_i$, we construct a label cost function $f_i: \mathsf{L} \to \R_{\geq 0}$.
The construction method of these $k$ functions is identical and thus we fix an index $i\in[k]$ and describe how to construct $f_i$ according to $c_i$.
For notation convenience, we use $c$ (resp. $f$) to denote $c_i$ (resp. $f_i$) in the following.

Given an arbitrary feasible label assignment $(l_{\bv})_{\bv\in\bV}$, we shall distribute the value of $f(\cL)$ among all nodes in $\bT$.
Formally, we will define a cost function $f_{\bv}:L_{\bv}\to\R_{\geq 0}$ for each node, and the value of function $f$ is just defined as the summation over all nodes, i.e., $f((l_{\bv})_{\bv\in\bV}):=\sum_{\bv\in\bV}f_{\bv}(l_{\bv})$.
The definition of the function $f$ directly follows if we know the definition of $f_{\bv}$ for any $\bv$.
In the following, we shall focus on how to define the cost function $f_{\bv}$.

By the completeness (\ref{prop:completeness}) of $\bT$, any edge $(a,b)\in E$ must be included in some node of $\bT$.
For each node $\bv\in\bV$, recall that $E_{\bv} \subseteq E$ is a set of edges such that, for each $(a,b)\in E_{\bv}$, node $\bv$ is the highest node that contains edge $(a,b)$.
Note that the highest node for each edge is unique.
This property is crucial which allows us to avoid redundant cost calculations for the same edge.
Before giving the formal definition, we introduce the following notation which characterizes a set of chosen edges in some node given a possible label.  
For each node $\bv \in \bV$ and one of its label $l_{\bv}\in L_{\bv}$, let $\edge(\bv,l_{\bv})\subseteq E_{\bv}$ be the set of edges such that (\rom{1}) each edge $(a,b)$ in $\edge(\bv,l_{\bv})$ is included in $E_{\bv}$;
(\rom{2}) for each edge $(a,b)\in\edge(\bv,l_{\bv})$, edge $(a,b)$ is chosen by $l_{\bv}$, i.e., $\chng(a,b)=\p$ in $l_{\bv}$.
We only care about these edges because once an edge in $\bv$ is chosen, such an edge would contribute some cost to the final value of the label assignment.
The formal definition of the cost function can be found in \cref{def:label-cost}.

\begin{definition}[Label Cost Function]
Let $\bv\in\bV$ be an arbitrary node in $\bT$ and $l_{\bv}\in L_{\bv}$ be a label of $\bv$.
Then, $f_{\bv}(l_{\bv}):=\sum_{e\in \edge(\bv,l_{\bv})}c(e)$ if $\edge(\bv,l_{\bv})\ne\emptyset$; otherwise, $f_{\bv}(l_{\bv}):=0$. 
\label{def:label-cost}
\end{definition}

\paragraph{Example.} 
Let us consider the example in \cref{fig:treewidth_reduction_new}.
Let $\bv$ be the node $sbdt$ in subfigure (\rom{2}) of \cref{fig:treewidth_reduction_new}.
Then, we have $E_{\bv}=\set{(s,b),(b,d)}$ and each edge in $E_{\bv}$ has a choosing label.
Since $\abs{\bv}=4$, $\bv$ have $12$ connectivity labels in total.
For notation convenience, let $e:=(s,b)$ and $e':=(b,d)$.
Recall that $c:2^{E}\to \R_{\geq 0}$ is the cost function of the original directed graph, and thus the cost of $e$ and $e'$ are $c(e)$ and $c(e')$, respectively.
Thus, when the choosing label of $e$ (resp. $e'$) is $\p$, it would have a contribution to $f_{\bv}$ which is the $c(e)$ (resp. $c(e')$).
Hence, for any label $l_{\bv}\in L_{\bv}$, $\edge(\bv,l_{\bv})$ must be the one of the following cases:
(\rom{1}) $\edge(\bv,l_{\bv})=\set{e}$, if the choosing label of $e$ is $\p$ and the choosing label of $e'$ is $\n$;
(\rom{2}) $\edge(\bv,l_{\bv})=\set{e'}$, if the choosing label of $e$ is $\n$ and the choosing label of $e'$ is $\p$;
(\rom{3}) $\edge(\bv,l_{\bv})=\set{e,e'}$ if the choosing labels of $e$ and $e'$ are $\p$.
Then, for an arbitrary label $l_{\bv}\in L_{\bv}$, we have:
$$
f_{\bv}(l_{\bv})=
\begin{cases}
c(e), & \text{if $\chng(e)=\p$ and $\chng(e')=\n$;}\\
c_(e'), &\text{if $\chng(e)=\n$ and $\chng(e')=\p$;} \\
c(e)+c(e'), & \text{if $\chng(e)=\p$ and $\chng(e')=\p$;}\\
0, & \text{otherwise.}
\end{cases}
$$

\cref{lem:reduction:cost} captures the equivalence of the cost function $c$ for the original instance and the cost function $f$ for the tree labeling instance.

\begin{lemma}
An $s \tto t$ path $\cP$ with cost $c(\cP)$ can be converted to a feasible label assignment $\cL$ such that $f(\cL)=c(\cP)$ and a feasible label assignment $\cL$ with cost $f(\cL)$ can be converted to a subgraph $G'$ containing at least one $s \tto t$ path such that $c(G')=f(\cL)$.
\label{lem:reduction:cost}
\end{lemma}

\begin{proof}
We show the lemma by proving the correctness of both directions.

\paragraph{From $\cP$ to $\cL$.}
Given an $s \tto t$ path $\cP$ of the original graph $G$ with the cost $c(\cP)$, we construct a feasible label assignment $\cL:=(l_{\bv})_{\bv\in\bV}$ such that $f(\cL) = c(\cP)$.
We use the same construction method stated in the proof of \cref{lem:reduction:map}.
By the proof of \cref{lem:reduction:map}, we know that the construction would give a feasible label assignment.
Now, we claim that $f(\cL)$ has the same value as $c(\cP)$.
To this end, we only need to show that $\set{\edge(\bv,l_{\bv})}_{\bv\in\bV}$ forms a partition of $\cP$.
Then, by the definition of the cost function (\cref{def:label-cost}), it is straightforward to see that $f(\cL)=c(\cP)$.
Observe that $\edge(\bv,l_{\bv}) \cap \edge(\bu,l_{\bu}) =\emptyset$ for any two nodes in $\bT$ since $E_{\bv}\cap E_{\bu}=\emptyset$.
Then, observe the following facts: (\rom{1}) for each edge $e\in\cP$, $e$ must be included in some $\edge(\bv,l_{\bv})$; (\rom{2}) $\bigcup_{\bv\in\bV}\edge(\bv,l_{\bv})$ can only include the edges from $\cP$.
Thus, $\set{\edge(\bv,l_{\bv})}_{\bv\in\bV}$ is a partition of $\cP$

\paragraph{From $\cL$ to $G'$.}
Given a feasible label assignment $\cL:=(l_{\bv})_{\bv\in\bV}$ with $f(\cL)$, we construct a subgraph $G'$ of the original graph such that (\rom{1}) $c(\cP) = f(\cL)$; (\rom{2}) $G'$ contains at least one $s \tto t$ path.
We use the same construction method stated in the proof of \cref{lem:reduction:map}.
Based on the proof of \cref{lem:reduction:map}, we know that the constructed subgraph must contain at least one $s \tto t$ path since $\cL$ is a feasible label assignment.
We claim that the cost of $G'$ is the same as the cost of the feasible label assignment.
Let $E'\subseteq E$ be the set of edges that are included in the subgraph $G'$.
The proof is the same as the previous direction, i.e., $\set{\edge(\bv,l_{\bv})}_{\bv\in\bV}$ is a partition of $E'$.
Then, by \cref{def:label-cost} we have $c(\cP)=f(\cL)$.
\end{proof}

\subsection{Proof of \cref{thm:ratio:treewidth}}
\label{sec:treewidth-proof}

\begin{observation}
Let $\gs^*$ be the guessing value at the beginning of the last round of the while-loop (lines \ref{line:binary-search-1}-\ref{line:binary-search-2}) in \cref{alg:treewidth}.
Then, we have $\gs^* \leq 2\cdot \opt$.
\label{obs:guess}
\end{observation}

\begin{proof}
Let $\lp$ be the minimum value of $\gs$ such that the constructed tree-labeling instance can have a valid label assignment.
Clearly, $\lp$ is a lower bound of the optimal solution, i.e., $\lp \leq \opt$.
Now we consider the last round of the while-loop, which implies that the current value of $\gs^*$ is still large enough to make the constructed tree-labeling instance admit a valid label assignment.
Since the current while-loop is the last round, we know that if we cut edges according to $\frac{\gs^*}{2}$, the constructed tree-labeling instance will not have a valid label assignment.
Thus, we have $\frac{\gs^*}{2}\leq \lp \leq\opt$.
\end{proof}

\begin{lemma}
Consider an arbitrary agent $i\in[k]$, \cref{alg:treewidth} finds a subgraph $G'$ such that (\rom{1}) $G'$ contains at least one $s \tto t$ path; (\rom{2}) $c_i(G') \leq 4H\cdot \log k \cdot \gs$, where $H$ is the height of the tree decomposition and $\gs$ is a guess of the optimal objective value such that the corresponding tree-labeling instance can have a valid label assignment.
\label{lem:treewidth:key}
\end{lemma}

\begin{proof}
Note that $\gs=1$ always holds since we do normalization for each $\gs$ (line \ref{line:cut-edge-2} of \cref{alg:treewidth}).
By \cref{lem:tree-labeling-ratio}, we know that there is a randomized algorithm that outputs a consistent label assignment $\cL$ such that for each $i\in[k]$, we have
$$
\E\left[\exp\left( \ln(1+\frac{1}{2H})\cdot f_i(\cL) \right)\right] \leq 1+\frac{1}{H}.
$$
Then, we have the following inequalities:
\begin{align*}
&\E\left[\left(1+\frac{1}{2H}\right)^{f_i(\cL)}\right] \leq 1+\frac{1}{H} \tag*{[By \cref{lem:tree-labeling-ratio}]} \\
\Rightarrow & \Pr\left[ \left(1+\frac{1}{2H}\right)^{f_i(\cL)} \geq k^2 \right] \leq \frac{1}{k^2}+\frac{1}{k^2 H} \tag*{[By Markov bound]} \\
\Rightarrow & \Pr\left[ f_i(\cL) \cdot \log(1+\frac{1}{2H}) \geq 2\log k \right] \leq \frac{1}{k^2}+\frac{1}{k^2 H} \\
\Rightarrow & \Pr\left[ f_i(\cL) \geq 8H\log k \right] \leq \frac{1}{k^2}+\frac{1}{k^2 H} \tag*{[By $\log(1+\frac{1}{2H}) \geq \frac{1}{4H}$]} \\
\Rightarrow & \Pr\left[ f_i(\cL) \geq 8H\log k \cdot \gs \right] \leq \frac{1}{k^2}+\frac{1}{k^2 H} \tag*{[By $\gs=1$]} \\
\end{align*}   
By \cref{lem:reduction:map}, we are able to convert such a consistent label assignment into a subgraph $G'$ such that $G'$ contains at least one $s \tto t$ path.
By \cref{lem:reduction:cost}, we know that $f_i(\cL)=c_i(G')$ for each $i\in[k]$.
Thus, for each $i\in[k]$, we have:
$$
\Pr\left[ c_i(G') \geq 8H\log k \cdot \gs \right] \leq \frac{1}{k^2}+\frac{1}{k^2 H} 
$$
\end{proof}

\begin{proofof}{\cref{thm:ratio:treewidth}}
After doing the doubling step, we can get a guess $\gs^*$ such that $\gs^* \leq 2 \cdot \opt$ by \cref{obs:guess}.
Since $\gs^*$ is also a guess of the optimal objective value such that the corresponding tree-labeling instance admits a valid label assignment, $\gs^*$ satisfies \cref{lem:treewidth:key}.
By \cref{lem:treewidth:key}, we know that there is a randomized algorithm that outputs a subgraph $G'$ containing an $s \tto t$ path such that, for each $i\in[k]$:
$$
\Pr\left[ c_i(G') \geq 8H\log k \cdot \gs^* \right] \leq \frac{1}{k^2}+\frac{1}{k^2 H}.
$$
By union bound, we have:
$$
\Pr\left[ \max_{i\in[k]}c_i(G') \leq 8H\log k \cdot \gs^* \right] \geq 1- \left(\frac{1}{k}+\frac{1}{k H}  \right).
$$
By \cref{lem:reduction:map}, we know that $G'$ contains at least one $s \tto t$ path and let $\cP$ be an arbitrary $s \tto t$ path in $G'$.
Since $\cP$ is only a part of $G'$, we have $c_i(\cP) \leq c_i(G')$ for each $i\in[k]$.
Thus, we have:
$$
\Pr\left[ \max_{i\in[k]}c_i(\cP) \leq \max_{i\in[k]}c_i(G') \leq 8H\log k \cdot \gs^* \right] \geq 1- \left(\frac{1}{k}+\frac{1}{k H}  \right).
$$
Note that $H$ is the height of the tree decomposition.
By \cref{lem:tree_decomp_height}, we know that $H\leq 2\ceil{\log_{\frac{5}{4}} 2n} \leq 8\log_2 n + 10$.
Plugging in, we have:
$$
\Pr\left[ \max_{i\in[k]}c_i(\cP) \leq (64\log n \log k+80\log k) \cdot \gs^* \right] \geq 1- \left(\frac{1}{k}+\frac{1}{k H}  \right).
$$
Note that $\gs^* \leq 2\cdot\opt$ and thus we conclude:
$$
\Pr\left[ \max_{i\in[k]}c_i(\cP) \leq (128\log n \log k+160\log k) \cdot \opt \right] \geq 1- \left(\frac{1}{k}+\frac{1}{k H}  \right).
$$

Now, we analyze the running time of \cref{alg:treewidth}.
It is not hard to see that solving the tree-labeling instance is the dominant step among the four steps of \cref{alg:treewidth}.
Thus, we only discuss the running time for this part.
By \cref{lem:treewidth:key}, we know that outputting a subgraph takes $\poly(n) \cdot \Delta^{O(H)}$ times, where $\Delta$ is the maximum size of label set among all nodes and $H$ is the height of the tree decomposition.
By our construction, we add $s$ and $t$ to each node $\bv$.
Thus, for each node $\bv$, we have at most $\ceil{\frac{\abs{\bv}+2}{2}}$ choosing labels since there are at most $\ceil{\frac{\abs{\bv}+2}{2}}$ edges contained in $\bv$, and we have at most $(\abs{\bv}+2)^2$ connectivity labels.
Thus, node $\bv$ has at most $(\abs{\bv}+2)^2+\frac{\abs{\bv}+2}{2}+1$ labels.
The original graph has treewidth at most $\ell$ and shrinking the height of the tree decomposition increases the treewidth to at most $3\ell+2$ by \cref{lem:tree_decomp_height}.
Thus, the width of the tree decomposition that we used is at most $3\ell + 4$ since we add $s$ and $t$ to each node.
This implies that $\abs{\bv}\leq 3\ell +5$ for all $\bv\in\bV$ since treewidth minus one is the size of the largest node.
Thus, any node in the tree decomposition that we used in \cref{alg:treewidth} has at most $(3\ell +7)^2+\frac{3\ell +7}{2}+1 \leq (3\ell + 8)^2$ labels.
Each label has two possibilities and thus we conclude that $\Delta \leq 2^{(3\ell+8)^2}$.
Recall that $H\leq 8\log n + 10$.
Thus, the running of solving tree-labeling instance is $\poly(n) \cdot n^{O(\ell^2)}$.
\end{proofof}

\section{Hardness of the Maximin Objective}
\label{sec:hardness}

In this section, we show that no approximation algorithm has a bounded approximation ratio for several classical problems under the maximin objective unless $\PP=\NP$.
Specifically, we show that the following problems are not approximable:
(\rom{1}) the robust $s \tto t$ path problem under the maximin objective (\cref{sec:hardness:stpath});
(\rom{2}) the robust weighted independent set problem on trees and interval graphs under the maximin objective (\cref{sec:hardness:is});
(\rom{3}) the robust spanning tree problem under the maximin objective (\cref{sec:hardness:sptree}).
Before presenting the proofs of these problems, we first show that the following variant of the set cover problem is NP-Complete (\cref{sec:hardness:sc}).
We shall use this hardness result to build the other reductions later.
We remark that these three reductions follow the same basic idea and thus they are similar.

\subsection{A Set Cover Problem}
\label{sec:hardness:sc}

\paragraph{The 2-choose-1 Set Cover Problem.}

Given a ground set of elements $U:=\set{u_1,\ldots,u_n}$, there are $\kappa$ collections $\cC:=\set{\cC_1,\ldots,\cC_{\kappa}}$ of subsets defined over $U$.
Each collection $\cC_i:=\set{S_p^i,S_q^i}$ consists of two subsets, where $S_p^i,S_q^i\subseteq U$.
A feasible solution $\cS\subseteq\cC$ is a subcollection such that $\cS$ contains exactly one subset from each $\cC_i,i\in[\kappa]$.
The goal is to determine whether there exists a feasible solution $\cS$ such that $\bigcup_{S\in\cS}S = U$.

We shall do the reduction from the classical 3-SAT problem.
In the following, we restate the problem for completeness.

\paragraph{The 3-SAT Problem.}
There is a set of binary variables $\set{x_1,\ldots,x_n}$ ($x_i\in\set{0,1}$ for all $i\in[n]$).
A literal is defined to be either a variable itself (e.g., $x_i=1$) or the negation of a variable (e.g., $\bar{x}_i=0$ since $x_i=1$).
Recall the OR ($\vee$) operator: $x_1\vee x_2 \vee \cdots \vee x_n=1$ if there exists an $i\in[n]$ such that $x_i=1$.
A clause is a set of three literals that connect with the OR operation.
Given an assignment of variables, a clause is said to be satisfied if its value is $1$.
Now, we are given $m$ clauses, and the goal is to determine whether there is an assignment such that all clauses can be satisfied.

\begin{lemma}
The 2-choose-1 set cover problem is NP-Complete. 
\label{lem:hardness:sc}
\end{lemma}

\begin{proof}
To simplify the proof, we show that a special case of 2-choose-1 set cover is NP-Complete.
In the special case, we assume that the collection $\cC$ has the following property:
each element $a_i\in U$ appears in exactly three sets in the collection $\cC$. 

We do the reduction from the 3-SAT problem.
For each $\cC_{i}:=\set{S_p^i,S_q^i}$, we create a binary variable $x_i$; choosing $S_p$ corresponds to setting $x_i=1$ and choosing $S_q$ corresponds to setting $\bar{x}_i=1$.
In total, we have $\kappa$ variables in the constructed 3-SAT instance.
Each element $a_i$ appears in exactly three sets, denoted by $S_f^i,S_h^i,S_g^i \in \cC$.
By the construction above, each set corresponds to a literal (either $x$ or $\bar{x}$), and thus we assume that $f_i,h_i,g_i$ are the corresponding literals of $S_f^i,S_h^i,S_g^i$.
For each element $a_i\in U$, we create a 3-CNF clause which consists of three literals $f_i,h_i,g_i$.
In total, we have $n$ clauses in the constructed 3-SAT instance.

By the above construction, a feasible solution that covers all elements in $U$ is equivalent to an assignment that satisfies all clauses, and vice versa.
Thus, the above special case of 2-choose-1 set cover is equivalent to 3-SAT.
\end{proof}

\subsection{Hardness of Maximin $s \tto t$ Path}
\label{sec:hardness:stpath}

In this section, we show a strong lower bound of the maximin $s \tto t$ path problem.
In the maximin $s \tto t$ path problem, we aim to find an $s \tto t$ path $\cP$ such that $\min_{i\in[k]}c_i(P)$ is minimized.
Technically, the cost function should be called the utility function in the current case, but we still use the name of the cost function to avoid misunderstanding.
It is well-known that finding the longest path on general graphs is an NP-Hard problem~\cite{schrijver2003combinatorial}.
But finding the longest path is equivalent to finding the shortest path when the input graph is a directed acyclic graph (DAG).
In the following, we show that the maximin $s \tto t$ path is not approximable even if the input graph is a series-parallel graph which is a special case of a DAG.

\begin{theorem}
Given any series-parallel graph and $k$ $\set{0,1}$-cost functions, no polynomial time algorithm is $\alpha$-approximate for the maximin $s \tto t$ path problem unless $\PP=\NP$, where $\alpha$ is an arbitrary function of the input.
\label{thm:hardness:maximin-st}
\end{theorem}

\begin{proof}
To show the theorem, it is sufficient to show the hardness of the following decision problem:
given any series-parallel graph and $k$ cost functions, determining whether the optimal solution is $0$.
Suppose that the above decision problem is NP-Complete.
Then, the optimization problem shall not admit any polynomial time algorithm with a bounded approximation ratio.
Otherwise, such an algorithm, denoted by $\alg$, can be used to determine whether the optimal solution is $0$.
Namely, the following statement is true: $\alg$ returns a solution with a zero objective value if and only if the optimal objective value is zero.
If $\alg$ returns a solution with a zero objective value, then the optimal solution must also have a zero objective value since $\alg \geq c\cdot \opt$ by the definition of the approximation ratio.
In another direction, if the optimal solution has a zero objective, then $\alg$ must return a solution with a zero objective value since $\alg \leq \opt$.
Thus, in the following, we focus on the hardness proof of the decision problem above.

We shall do the reduction from the 2-choose-1 set cover problem.
For each collection $\cC_i:=\set{S_p^i,S_q^i}$, we first construct a directed graph with four vertices $\set{s_i,S_p^i,t_i,S_q^i}$ and four edges $(s_i,S_p^i)$, $(s_i,S_q^i)$, $(S_p^i,t_i)$ and $(S_q^i,t_i)$.
We refer to such a directed graph as a {\em meta-graph}.
See \cref{fig:maximin-st-path} for an example.
For each $\cC_i$, we create a meta-graph.
And then, we create a source vertex $s$ and sink vertex $t$ and connect them with all meta-graphs by directed edges: $(s,s_1)$, $(t_1,s_2)$, $\ldots$ and $(t_{\kappa},t)$.
An example can be found in \cref{fig:maximin-st-path}.
Let $G:=(V,E)$ be the constructed series-parallel graph.
For each element $u_i\in U$, we create an agent $a_i$ with a cost function $c_i$.
The cost function $c_i$ is defined as: (\rom{1}) for all $j\in[\kappa]$, $c_i(s_j,S_p^j)=1$ if $u_i\in S_p^j$; 
(\rom{2}) for all $j\in[\kappa]$, $c_i(s_j,S_q^j)=1$ if $u_i\in S_p^j$; (\rom{3}) $c_i(e)=0$ for all other edges $e$.

\begin{figure}[htb]
    \centering
    \includegraphics[width=15cm]{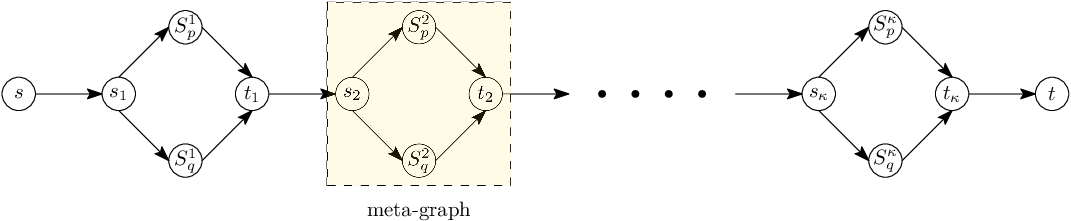}
    \caption{The constructed series-parallel graph according to the 2-choose-1 set cover instance.}
    \label{fig:maximin-st-path}
\end{figure}

By the above construction, an $s \tto t$ path of $G$ with a non-zero objective value shall imply a feasible solution that covers all elements.
And a feasible solution that can cover all elements shall imply an $s \tto t$ path with non-zero objective value.
Thus, a polynomial algorithm that can determine whether the given maximin $s \tto t$ path instance has a non-zero optimal objective value shall imply a polynomial time algorithm that is able to determine whether the constructed 2-choose-1 set cover instance has a feasible solution that covers all elements.
Thus, the decision problem is NP-Complete.
\end{proof}

\subsection{Hardness of Maximin Weighted Independent Set}
\label{sec:hardness:is}

In the section, we show that the problem of maximin weighted independent set is not approximable even if the input graph is an interval graph or tree.
This provides a strong lower bound for this problem.
The previous works~\cite{DBLP:journals/symmetry/KlobucarM21,DBLP:journals/ol/NobibonL14} only show the NP-Hardness of this problem.

\paragraph{Maximin Weighted Independent Set.}
Given an undirected graph $G:=(V,E)$, there are $k$ additive weight functions $c_1,\ldots,c_k$.
Each weight function $c_i:2^{V}\to\R_{\geq 0}$ assigns a non-negative weight to each vertex in $V$.
The goal is to compute an independent set $S\subseteq V$ such that $\min_{i\in[k]}c_i(S)$ is maximized.

\begin{theorem}
Given any tree or interval graph and $k$ $\set{0,1}$-weight functions, no polynomial time algorithm is $\alpha$-approximate for the maximin weighted independent set problem unless $\PP=\NP$, where $\alpha$ is an arbitrary function of the input.
\label{thm:hardness:maximin-is}
\end{theorem}

\begin{proof}

To show the theorem, it is sufficient to show the hardness of the following decision problem by the same reason stated in the proof of \cref{thm:hardness:maximin-st}:
given any tree or interval graph, determining whether the optimal solution is $0$.
For both problems, we shall do the reduction from the 2-choose-1 set cover problem.

We start with the easier reduction: the maximin weighted independent set on interval graphs.
For each collection $\cC_i:=\set{S_p^i,S_q^i}$, we first construct an undirected graph with only two vertices $\set{S_p^i,S_q^i}$ and one edge between these two vertices.
We refer to such an undirected graph as a {\em meta-graph}. 
For each $\cC_i$, we create a meta-graph.
And then, the final graph $G:=(V,E)$ just consists of these meta-graphs.
Note that there is no edge between any meta-graphs, i.e., $G$ consists of $\kappa$ connected components and each connected component is just an edge.
It is easy to see that the constructed undirected graph $G$ is an interval graph.
For each element $u_i\in U$, we create an agent $a_i$ with a weight function as follows:
(\rom{1}) for all $j\in[\kappa]$, $c_i(S_p^j)=1$ if $u_i\in S_p^j$;
(\rom{2}) for all $j\in[\kappa]$,
$c_i(S_q^j)=1$ if $u_i\in[\kappa]$;
(\rom{3}) $c_i(v)=0$ for all other vertices.

Now we show the hardness of the tree by slightly modifying the above reduction.
For each collection $\cC_i:=(S_p^i,S_q^i)$, we first construct an undirected graph with three vertices $v_i,S_p^i,S_q^i$ and two edges $(v_i,S_q^i)$, $S_p^i,S_q^i$.
We refer to such an undirected graph as a meta-graph.
For each $\cC_i$, we create a meta-graph, and then, we connect these meta-graphs via edges $(S_q^1,v_2)$, $(S_q^2,v_3)$, $\ldots$, $(S_{q}^{\kappa-1},v_{\kappa})$.
An example can be found in \cref{fig:maximin-wis-tree}.
It is easy to see that the constructed graph is a tree.
We then use the same cost function as the reduction for the interval graph.

\begin{figure}[htb]
    \centering
    \includegraphics[width=15cm]{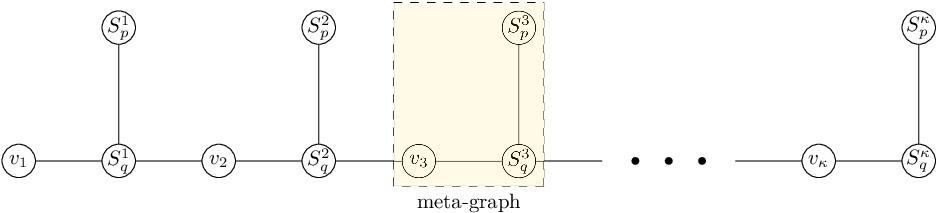}
    \caption{The constructed tree according to the 2-choose-1 set cover instance.}
    \label{fig:maximin-wis-tree}
\end{figure}    

By the above constructions, an independent set with a non-zero objective shall imply a feasible solution that covers all elements.
And a feasible solution that can cover all elements shall imply an independent set with non-zero objective value.

\end{proof}

\subsection{Hardness of Maximin Spanning Tree}
\label{sec:hardness:sptree}

This section shows that the maximin spanning tree problem is not approximable.
This is quite different from the minimax spanning tree problem, which admits a $O(\frac{\log k}{\log \log k})$-approximation algorithm~\cite{DBLP:conf/focs/ChekuriVZ10}.
And this result is almost tight for the minimax spanning tree by the lower bound of $\Omega(\log^{1-\epsilon} k)$ stated in~\cite{DBLP:journals/tcs/KasperskiZ11}.
We shall do the reduction from another variant of the sect cover problem called {\em the 3-choose-2 set cover problem}.
The NP-Completeness of this problem is built from the hardness of the 2-choose-1 set cover problem.

\paragraph{Maximin Spanning Tree.}

Given an undirected graph $G:=(V,E)$, there are $k$ additive weight functions $c_1,\ldots,c_k$ defined over edges.
Each weight function $c_i:2^{E}\to\R_{\geq 0}$ assigns a non-negative weight to each edge.
The goal is to find a spanning tree $T\subseteq E$ such that $\min_{i\in[k]}c_i(T)$ is maximized.

\paragraph{The 3-choose-2 Set Cover Problem.}

Given a ground set of elements $U:=\set{u_1,\ldots,u_n}$, there are $\kappa$ collections $\cC:=\set{\cC_{1},\ldots,\cC_{\kappa}}$ of subsets defined over $U$.
Each collection $\cC_i$ consists of three subsets, i.e., $\cC_i:=\set{S_a^i,S_b^i,S_c^i}$, where $S_a^i,S_b^i,S_c^i \subseteq U$.
A feasible solution $\cS\subseteq \cC$ is a subcollection such that $\cS$ contains exactly two subsets from each $\cC_i,i\in[\kappa]$.
The goal is to determine whether there exists a feasible solution $\cS$ such that $\bigcup_{S\in\cS}S=U$.

\begin{corollary}
The 3-choose-2 set cover problem is NP-Complete.
\end{corollary}

\begin{proof}
The proof is straightforward from the 2-choose-1 set cover problem by adding $\kappa$ dummy elements and $\kappa$ dummy sets.
Formally, given any 2-choose-1 set cover instance, we add dummy elements $u_1^*,\ldots,u_{\kappa}^*$ to $U$ and add a dummy subset $S_{i}^*:=\set{u_i^*}$ to $\cC_{i}$ for each $i\in[\kappa]$.
In this way, if a feasible solution $\cS$ to 3-choose-2 set cover instance covers all ground elements, then $\cS$ must choose $S_i^*$ from each $\cC_i$.
Thus, it shall imply a feasible solution to the 2-choose-1 set cover instance that covers all ground elements.
In another direction, if there is a feasible solution to the 2-choose-1 set cover instance that covers all elements, then such a feasible solution can also cover all ground elements by choosing $S_i^*$ from each $\cC_i$.
\end{proof}

\begin{theorem}
Given any undirected graph and $k$ $\set{0,1}$-weight functions, no polynomial time algorithm is $\alpha$-approximate for the maximin spanning tree problem unless $\PP=\NP$, where $\alpha$ is an arbitrary function of the input.
\label{thm:hardness:maximin-stree}
\end{theorem}

\begin{proof}
    
Same as the previous proofs, we show that the following decision problem is NP-Complete:
given any undirected graph and $k$ weight functions, determining whether the optimal solution is $0$.

We shall do the reduction from the 3-choose-2 set cover problem.
For each collection $\cC_i:={S_a^i,S_b^i,S_c^i}$, we construct an undirected triangle with three vertices $v_a^i,v_b^i,v_c^i$ and three edges $S_a^i,S_b^i,S_c^i$.
As usual, we refer to such a triangle as a meta-graph.
For each collection $\cC_i$, we create a meta-graph, and then, we connect them by edges $(v_c^1,v_a^2),\ldots,(v_c^{\kappa-1},v_a^{\kappa})$.
An example can be found in \cref{fig:maxmin-stree}.
For each element $u_i\in U$, we create an agent with weight function $c_i$ as follows:
(\rom{1}) for all $j\in[\kappa]$, $c_i(S_a^j)=1$ if $u_i\in S_a^j$;
(\rom{2}) for all $j\in[\kappa]$, $c_i(S_b^j)=1$ if $u_i\in S_b^j$;
(\rom{3}) for all $j\in[\kappa]$, $c_i(S_c^j)=1$ if $u_i\in S_c^j$;
(\rom{4}) $c_i(e)=0$ for all other edges. 

\begin{figure}[htb]
    \centering
    \includegraphics[width=15cm]{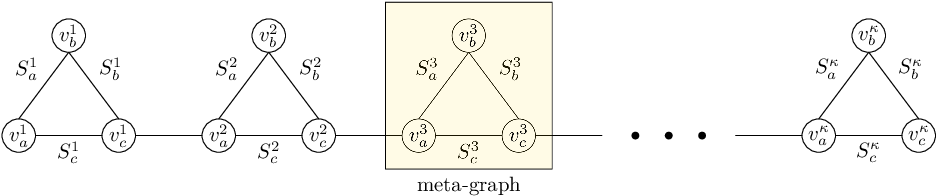}
    \caption{The constructed graph according to the 3-choose-2 set cover instance.}
    \label{fig:maxmin-stree}
\end{figure}

By the above construction, a spanning tree with a non-zero objective shall imply a feasible solution that covers all elements, and vice versa.

\end{proof}

\section{Conclusion}

In this work, we studied the robust $s \tto t$ path problem and designed algorithms with polylogarithmic approximation ratios on different graph classes.
For the class of graphs with treewidth at most $\ell$, we obtained a $O(\log n\log k)$-approximate algorithm whose running time is a function of $\ell$.
This implies a polynomial time algorithm for graphs with bounded treewidth.
This result partially answers the open question in~\cite{DBLP:journals/corr/abs-1806-08936} in which they ask whether it is possible to remove $\sqrt{n}$ from the approximation ratio.
For series-parallel graphs, we showed that there is an LP-based randomized rounding algorithm that is $O(H\log k)$-approximate.
When the decomposition tree of the series-parallel graph has a polylogarithmic depth, the algorithm achieves a polylogarithmic ratio.
For general graphs, we proved that there is a quasipolynomial time algorithm that is $O(\log n \log k)$-approximate.
Our approach is based on a novel linear program that enables us to get rid of the $\Omega(\max\{k,\sqrt{n}\})$ integrality gap from the natural linear program.
We also investigated the robustness of the shortest path problem under the maximin criteria and showed that the problem is not approximable.
Our reduction also works for other robust optimization problems under the maximin objective, such as maximin weighted independent set on trees or interval graphs, and maximin spanning tree.
And thus, they are also not approximable.

An immediate direction is to show whether there is a polynomial time algorithm with a polylogarithmic approximation ratio for general graphs.
This includes two directions: either improve the running time of our algorithm (e.g., construct a supertree with polynomial size) or improve the lower bound of the problem.
Another interesting direction is to investigate whether there is a quasipolynomial time algorithm with a polylogarithmic approximation ratio for robust perfect matching under the minimax objective.
The best approximation ratio so far for robust matching is still $O(k)$, which a trivial algorithm can achieve.
It is well-known that the shortest path problem is a special case of min-cost perfect matching on the bipartite graph, and thus robust perfect matching generalizes robust $s \tto t$ path.
Hence, it would be interesting to show whether our methods can be extended to the robust matching problem.

\section*{Acknowledgment}
We thank the anonymous reviewers for their many insightful comments and suggestions.
Chenyang Xu was supported in part by Science and Technology Innovation 2030 –``The Next Generation of Artificial Intelligence" Major Project No.2018AAA0100900.
Ruilong Zhang was supported by NSF grant CCF-1844890.

\newpage
\clearpage
\bibliographystyle{plain}
\bibliography{main}

\newpage
\appendix
\section{Hard Instance for Dijkstra-type Algorithm}
\label{app:hard-greedy}

\begin{figure}[htb]
    \centering
    \includegraphics[width=13cm]{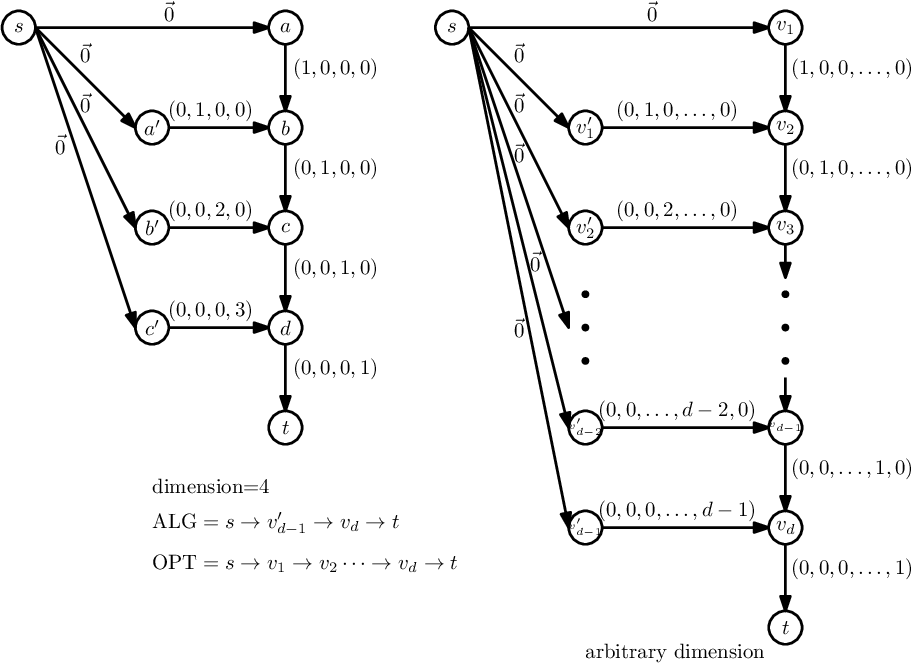}
    \caption{The hard instance for the Dijkstra-type algorithm. The left-hand side is an instance consisting of $4$ agents. The right-hand side is the general instance consisting of $d$ agent. The goal is to pick an $s$-$t$ path such that the $\ell_p$-norm cost is minimized overall agents. The edge cost of agents is marked as a vector on each edge. It can be seen that the Dijkstra-type algorithm shall choose the path $s\to v'_{d-1}\to t$ and its cost is $d$. The optimal solution will choose the path $s\to v_1 \to v_2 \to \cdots \to v_d \to t$ and its cost is $d^{1/p}$. Thus, the approximation ratio is at least $d^{1-1/p}$. When $p=O(\log d)$, the approximation ratio is $\Omega(d)$.}
    \label{fig:hard-greedy}
\end{figure}

The proposed algorithm~\cite{DBLP:conf/icalp/BiloC0FM17}(Algorithm 2) is the classical Dijkstra algorithm by replacing the distance with the $\ell_p$-norm. 
The hard instance for the algorithm is shown in~\cref{fig:hard-greedy}.
In the following, we briefly discuss how the algorithm performs on the constructed instance consisting of $4$ agents (e.g., the number of dimensions is $4$); this specific instance is shown on the left-hand side of \cref{fig:hard-greedy}.

Our constructed directed graph is a DAG, so \cite{DBLP:conf/icalp/BiloC0FM17}(Algorithm 2) becomes a dynamic programming-type algorithm, with nodes processed using a topological order. Namely, the algorithm stores the best path in the $\ell_p$-norm for every node. 
Note that the nodes $a, a', b', c'$, and $d'$ have path costs $(0, 0, 0, 0)$. 
Now, $b$ has two incoming edges, one from a with cost $(1, 0, 0, 0)$, and the other from $a'$ with cost $(0, 1, 0, 0)$. 
The two options have the same cost as they are symmetric. 
Let us say we choose the path from $a'$. So, the path cost for b will be $(0, 1, 0, 0)$. 
Now, we consider the node $c$. \
The path via $b$ to $c$ will have cost $(0, 2, 0, 0)$, and the path via $b'$ to $c$ will have cost $(0, 0, 2, 0)$. 
Again, they are symmetric, and assume we choose the latter option. 
Then, the path cost for $c$ will be $(0, 0, 2, 0)$. 
Keep going like this: the path cost for $d$ will be $(0, 0, 0, 3)$, and the path cost for $t$ will be $(0, 0, 0, 4)$. The path chosen by the algorithm is $s\to c' \to d \to t$, which has a cost $(0, 0, 0, 4)$.  
It can be seen that the optimum path is $s\to a \to b \to c \to d \to t$, which has cost $(1, 1, 1, 1)$. 
Thus, the algorithm’s cost is $4$ while the optimum cost is $4^{1/p}$.
Although the above hard instance depends on how the ties are broken in the algorithm, one can easily change the costs slightly so that there are no ties when we run the algorithm.

\section{Missing Materials from \cref{sec:sp}}

\subsection{Discussion of \eqref{sp:tree-LP}}
\label{sec:tree-lp:discussion}

The constraints \eqref{sp:tree-LPC:root}, \eqref{sp:tree-LPC:series} and \eqref{sp:tree-LPC:parallel} are used to ensure that an integral solution must be a feasible subtree.
\eqref{sp:tree-LPC:cost} ensures that the cost of the selected subtree is at most $1$ for all agents.
As one may observe, the natural formulation of the cost constraint is as follows:
$$
\sum_{\bu \in \bV} \xu \cdot f_i(\bu) \leq 1, \quad \forall i\in[k].
$$
This constraint is equivalent to the following constraint as $\xr=1$ by \eqref{sp:tree-LPC:root} of \eqref{sp:tree-LP}.
$$
\sum_{\bu \in \Lambda(\br)} \xu \cdot f_i(\bu) \leq \xr, \quad \forall i\in[k].   
$$
The above constraint is a natural cost constraint requiring that the whole selected subtree cost is at most $1$ for all agents.
Thus, one might expect that the following linear program \eqref{weak-tree-LP} is also a candidate, which replaces \eqref{sp:tree-LPC:cost} of \eqref{sp:tree-LP} with the constraint above.

\begin{align}
    &&  & \tag{\text{Weak-Tree-LP}} \label{weak-tree-LP}\\
    &&\sum_{\bu \in \Lambda(\br)} \xu \cdot f_i(\bu) &\leq \xr, & \forall i\in[k]\label{weak-tree-LPC:cost} \\
    &&\xr &=1 , & \label{weak-tree-LPC:root}\\
    &&\sum_{\bu\in\kid(\bv)}\xu &= \xv, &\forall \bv\in P(\bT) \label{weak-tree-LPC:parallel}\\
    &&\xu &= \xv, & \forall \bv\in S(\bT), \bu\in\kid(\bv) \label{weak-tree-LPC:series}\\
    &&\xv &\geq 0, &\forall \bv\in \bV 
\end{align}

It is easy to see that \eqref{sp:tree-LP} enhances \eqref{weak-tree-LP} since \eqref{sp:tree-LPC:cost} implies \eqref{weak-tree-LPC:cost}.
In fact, as one may observe, \eqref{weak-tree-LP} is equivalent to \eqref{En-Flow-LP} because \eqref{weak-tree-LP} does not utilize the tree structure.
The \eqref{sp:tree-LPC:root}, \eqref{sp:tree-LPC:series} and \eqref{sp:tree-LPC:parallel} of \eqref{weak-tree-LP} only ensure that the selected subtree corresponds to an s-t path; the role of these three constraints are equivalent to the constraints \eqref{flow-LPC:st} and \eqref{flow-LPC:equal} of \eqref{En-Flow-LP}.
Thus, \eqref{weak-tree-LP} also cannot overcome the hard instance in \cref{fig:flow_lp_gap}.
Formally, we have the following lemma (\cref{lem:weak-tree-lp-gap}).

\begin{lemma}
The integrality gap of \eqref{weak-tree-LP} is $\Omega (k)$.    
\label{lem:weak-tree-lp-gap}
\end{lemma}

\begin{figure}[htb]
    \centering
    \includegraphics[width=15cm]{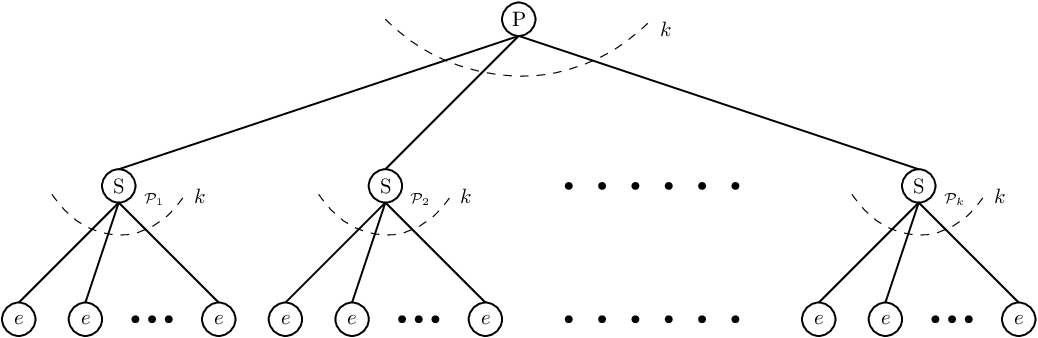}
    \caption{The decomposition tree of the series-parallel graph in \cref{fig:flow_lp_gap}.}
    \label{fig:weak-tree-lp-gap}
\end{figure}

\begin{proof}
The hard instance is the same as the hard instance for \eqref{En-Flow-LP} stated in \cref{fig:flow_lp_gap}.
Now, we draw the decomposition tree of the hard instance in \cref{fig:weak-tree-lp-gap}.
Let $\bT:=(\bV,\bE)$ be the decomposition tree of the series-parallel graph $G:=(V,E)$ in \cref{fig:flow_lp_gap}.
Note that $\bT$ only consists of three levels; the root is a parallel node, the middle level contains all series nodes, and each leaf nodes correspond to each edge in $E$.
Let $\leaf(\bT)$ be the set of leaf nodes in $\bT$.
Let $\leaf_i(\bT)$ be the set of leaf nodes whose parent node is the $i$-th series node.
Fix an arbitrary agent $i\in[k]$, by the definition of our cost function, we have (\rom{1}) $f_i(\bv)=1$ if $\bv\in \leaf_i(\bT)$; (\rom{2}) $f_i(\bv)=0$ if $\bv\notin \leaf_i(\bT)$. 

If the guessing value of the optimal solution is $1$, then same as the optimal solution to \eqref{En-Flow-LP}, the optimal solution $x^*:=(\xv)_{\bv\in\bV}$ to \eqref{weak-tree-LP} on such an instance shall have the following assignment:
(\rom{1}) $\xv=\frac{1}{k}$ for all $\bv\in\leaf(\bT)$;
(\rom{2}) $\xv=\frac{1}{k}$ for all $\bv\in S(\bT)$;
(\rom{3}) $\xr=1$ for the root of $\bT$.
Thus, \eqref{weak-tree-LP} can still have a feasible solution in this case, but the optimal integral solution is $1$ which leads to an integrality gap of $\Omega(k)$.
\end{proof}

However, the above hard instance can be fixed by \eqref{sp:tree-LP}.
Suppose that the guess value of the optimal solution is $\gs$, where $1\leq \gs<k$.
We show that \eqref{sp:tree-LP} cannot have any feasible solution in this case.
Note that if $\gs <1$, then all edges would be discarded in which the above claim trivially holds.
In the case where $1\leq \gs <k$, no edge will be discarded.
Suppose that the value of the $i$-th series node $\bv$ is $\xv\in[0,1]$.
By \eqref{sp:tree-LPC:series}, we know that the value of $\bv$'s child node is also $\xv$.
By \eqref{sp:tree-LPC:cost}, we have $k\cdot \xv \cdot \frac{1}{\gs} \leq \xv$ which implies that $\gs \geq k$ must hold.
Thus, \eqref{sp:tree-LP} cannot have any feasible solution as long as the guess is smaller than $k$ which is the optimal integral solution.

\subsection{Lower Bound Instance in~\cite{DBLP:journals/ipl/KasperskiZ09}}
\label{sec:sp:hard-instance}

In this section, we show that \cref{alg:sp-graph} is able to output an $O(\log \log n \log k)$-approximate solution for the hard instance stated in~\cite{DBLP:journals/ipl/KasperskiZ09} which leads to a lower bound of $\Omega(\log^{1-\epsilon} k)$ for any $\epsilon>0$.
Let $\cI:=(G,\cF)$ be such an instance, where $G$ is the directed graph in $\cI$ and $\cF$ is a set of cost functions with the size of $k$.
We formally show the following lemma (\cref{lem:lower-bound-instance}).

\begin{lemma}
Taking the instance $\cI=(G,\cF)$ as the input, \cref{alg:sp-graph} is an $O(\log \log n \log k)$-approximate algorithm, where $n$ is the number of vertices in $G$ and $k$ is the number of cost functions.    
\label{lem:lower-bound-instance}
\end{lemma}

By \cref{thm:sp-graph}, we know that the approximation ratio of \cref{alg:sp-graph} depends on the height of the decomposition tree of the series-parallel graph.
Thus, in the following, we can temporarily ignore the cost functions and focus on the decomposition tree of $G$.
Observe that if we can show that the series-parallel graph's decomposition tree in the hard instance~\cite{DBLP:journals/ipl/KasperskiZ09} is $O(\log \log n)$, then \cref{lem:lower-bound-instance} directly follows.
Thus, we only restate the definition of $G$ in the following.

\begin{figure}[htb]
    \centering
    \includegraphics[width=15cm]{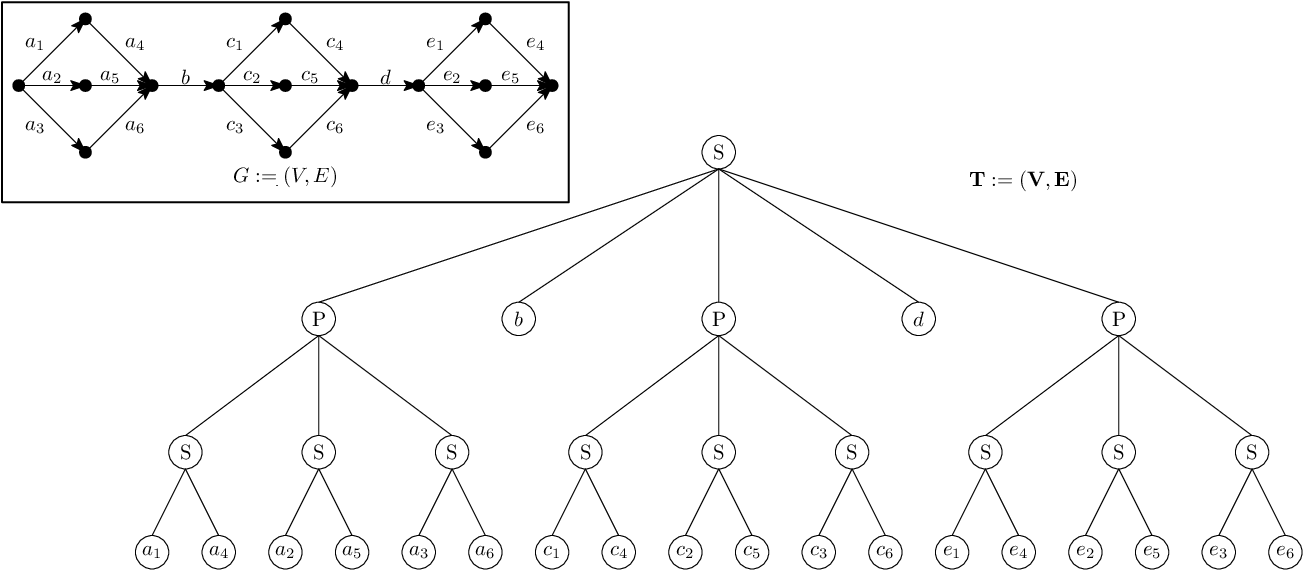}
    \caption{The hard instance leading to a lower bound of $\Omega(\log^{1-\epsilon}k)$ stated in~\cite{DBLP:journals/ipl/KasperskiZ09}. The figure only shows $G^{(1)}$ and its decomposition tree $\bT^{(1)}$. The whole instance requires $t$ times replacements, denoted by $G^{(t)}$ and $\bT^{(t)}$, where $t:=\log \log^{\beta} n$ for some constant $\beta>0$.}
    \label{fig:lower-bound-instance}
\end{figure}

The final series-parallel graph $G:=(V,E)$ is generated by a meta-graph via several replacements.
Thus, we shall restate the instance starting from the meta-graph, denoted by $G^{(1)}:=(V^{(1),E^{(1)}})$.
We slight abuse the notation and assume that $\abs{V^{(1)}}:=n$.
The meta-graph is merged from three smaller graphs with two edges via the series composition.
Each small graph just consists of $6$ edges from $s$ to $t$.
An example can be found in \cref{fig:lower-bound-instance}.
Then, the graph $G^{(2)}:=(V^{(1)},E^{(1)})$ is generated from $G^{(1)}$ by replacing every arc in $G^{(1)}$ by the whole graph $G^{(1)}$.
Similarly, the graph $G^{(p)}:=(V^{(p)},E^{(p)})$ is generated from $G^{(p-1)}$ by replacing every arc in $G^{(p-1)}$ by the whole graph $G^{(1)}$.
By repeating the above replacement $t$ time, we obtain the final graph $G^{(t)}$, where $t:=\log \log^{\beta} n$ and $\beta$ is some constant.
In~\cite{DBLP:journals/ipl/KasperskiZ09}, they actually proved the following statement (\cref{lem:log-lower-bound}).
From \cref{lem:log-lower-bound}, we can find that $\beta$ is some factor related to $\epsilon$, but it is a constant.

\begin{lemma}[\cite{DBLP:journals/ipl/KasperskiZ09}]
The robust $s \tto t$ path problem on the series-parallel graph cannot be approximated within a factor of $\log^{1-\epsilon} k$ for any $\epsilon >1-\frac{\beta}{\beta+1}$ unless $\NP \subseteq \mathrm{DTIME}(n^{\mathrm{poly}\log n})$.
\label{lem:log-lower-bound}
\end{lemma}

To obtain the decomposition tree of $G^{(t)}$, we can also simulate the process above, i.e., $\bT^{(t)}$ can be constructed from $\bT^{(1)}$ via $t$ times replacements.
For example, the decomposition tree $\bT^{(2)}$ of $G^{(2)}$ can be constructed by replacing every leaf node in $\bT^{(1)}$ by the whole tree $\bT^{(1)}$.
After that, we need to merge the adjacent series node to make sure that the generated tree is a decomposition tree.
Thus, the height of $\bT^{(2)}$ is the height of $\bT^{(1)}$ plus 2.
This implies that the height of $\bT^{(p)}$ is equal to the height of $\bT^{(p-1)}$ plus 2.
Thus, the height of $\bT^{(t)}$ is the height of $\bT^{(1)}$ plus $2\cdot t$, i.e., $4+2\log\log^{\beta} n=4+2\beta\log\log n$.
Note that the number of vertices in $G^{(t)}$ is roughly $N:=n^{\log\log^{\beta}n}$ and thus $4+2\beta \log \log N$ is $4+2\beta\log\log n+2\beta \log (\beta\log \log n)$.
Thus, the height of $\bT^{(t)}$ is at most $4+2\beta\log\log N$, where $N$ is the number of vertices in $G^{(t)}$.
Thus, \cref{alg:sp-graph} is an $O(\log \log n\log k)$-approximate algorithm.

\end{document}